\documentclass[12pt,a4paper]{article}
\usepackage[dvips]{graphicx}
\usepackage{subfigure}
\usepackage{amssymb,amsmath}
\usepackage{amsfonts}
\usepackage[mathscr]{eucal}
\usepackage{ascmac}
\usepackage{cite}
\usepackage{bm}
\usepackage[charter]{mathdesign}

\usepackage[height=22.5cm,width=16.5cm]{geometry}

\usepackage{multicol}
\usepackage[usenames,dvipsnames]{xcolor}
\definecolor{rosso}{cmyk}{0.1,0.85,0.9,0.3}
\definecolor{verde}{cmyk}{0.8,0,0.6,0.25}
\definecolor{bluc}{cmyk}{0.95,0.67,0,0}
\definecolor{blucc}{cmyk}{0.75,0.5,0,0}
\definecolor{tang}{cmyk}{0.01,0.44,0.98,0.03}

\newcommand{\nn}{\nonumber \\}
\newcommand{\bea}{\begin{eqnarray}}
\newcommand{\eea}{\end{eqnarray}}

\makeatletter
\@addtoreset{equation}{section}

\makeatother

\def\be{\begin{equation}}
\def\ee{\end{equation}}
\def\fr{\frac}
\def\der{\partial}

\def\({\left(}
\def\){\right)}
\def\1{^{(1)}}
\def\2{^{(2)}}
\def\a{^{(a)}}
\def\b{^{(b)}}

\def\cL{\mathcal{ L}}

\def\Tr{\mathrm{Tr}}

\begin{document}

\begin{titlepage}

\begin{flushright}
UT-13-30\\
\end{flushright}

\vskip 4cm

\begin{center}

{\LARGE \bfseries
Fate of $\bm{Z_2}$ Symmetric Scalar Field
}

\vskip .65in

{
Kyohei Mukaida$^{\spadesuit}$,
Kazunori Nakayama$^{\spadesuit,\diamondsuit}$
and Masahiro Takimoto$^{\spadesuit}$
}

\vskip .4in

\begin{tabular}{ll}
$^{\spadesuit}$ & \!\! {\em Department of Physics, Faculty of Science, }\\
& {\em University of Tokyo,  Bunkyo-ku, Tokyo 133-0033, Japan}\\[.5em]
$^{\diamondsuit}$ &\!\! {\em Kavli Institute for the Physics and Mathematics of the Universe, }\\
&{\em University of Tokyo,  Kashiwa, Chiba 277-8583, Japan}
\end{tabular}
\end{center}

\vskip .65in

\begin{abstract}
The evolution of a coherently oscillating scalar field with $Z_2$ symmetry is studied in detail.
We calculate the dissipation rate of the scalar field based on the closed time path formalism.
Consequently, it is shown that the energy density of the coherent oscillation can be efficiently dissipated
if the coupling constant is larger than the critical value,
even though the scalar particle is stable due to the $Z_2$ symmetry.

\end{abstract}

\end{titlepage}

\tableofcontents

\setcounter{page}{1}

\section{Introduction} \label{sec:introduction}

Scalar fields often appear in extensions of the Standard Model (SM) and play various roles in cosmology.
The impacts of a scalar field are most prominent if it obtains a large field value during inflation.
This is because the coherent oscillation of a scalar field has huge energy density, which might alter the subsequent
cosmological evolution scenarios.
In particular, the existence of such a scalar field with too long lifetime to decay before the big-bang nucleosynthesis (BBN)
is problematic. It is known as the cosmological moduli problem~\cite{Coughlan:1983ci,deCarlos:1993jw}.

Despite its importance, the dynamics of scalar fields in the hot thermal Universe has not been understood well.
Let us summarize current understandings on the fate of the scalar field coherent oscillation:
\begin{itemize}
\item In the limit of large scalar mass, the perturbative decay rate determines the epoch when it decays.

\item If the temperature of the cosmic plasma is (much) higher than the scalar field mass, the perturbative decay is blocked
due to thermal masses of the decay products.
Instead, the effects of thermal dissipation on the scalar field becomes important~\cite{Berera:1995ie,Yokoyama:2004pf,Drewes:2010pf}.

\item If the amplitude of the scalar field is large enough, it acts as the non-adiabatic background for the coupled particles.
It induces the non-perturbative particle production~\cite{Kofman:1994rk}.

\end{itemize}
All these effects significantly affect and complicate the scalar dynamics.
The full analysis in broad parameter ranges was carried out in Ref.~\cite{Mukaida:2012qn,Mukaida:2012bz}
in the case where the scalar field has a yukawa/gauge coupling to light fermions/bosons.

In this paper, we study the scalar field dynamics when the real scalar field $\phi$ has a $Z_2$ symmetry ($\phi \to -\phi$),
which is broken neither explicitly nor spontaneously.
Phenomenologically such a scalar field may be introduced to account for the dark matter (DM) of the Universe~\cite{Silveira:1985rk}.
Depending on its mass and couplings, its thermal relic abundance can be consistent with observed DM abundance.
However, this argument is based on an assumption that there is no coherent oscillation contribution.\footnote{
	It is possible that the scalar $\phi$ sits at the origin during inflation due to the positive Hubble induced mass term.
	In this case, the coherent oscillation is not induced.
}
Moreover, such a scalar field may play a role of inflaton or curvaton~(see \textit{e.g.} Refs.~\cite{Okada:2010jd,Enqvist:2012tc}).
One might expect that the scalar field is stable and hence the coherent oscillation of such a scalar field is cosmologically disastrous.
However, this intuition is not always true.
Although the perturbative decay around the vacuum $\phi = 0$ does not occur, 
the scalar field can dissipate its energy through scatterings with particles in thermal bath.
If this is efficient, the amplitude of the scalar field may damp to a cosmologically harmless level.

To be concrete, let us consider a following toy model invariant under $Z_2$: $\phi \mapsto -\phi$
\begin{align}
	\cL = \cL_\text{kin} - \frac{1}{2} m_\phi^2 \phi^2 - \lambda^2 \phi^2 |\chi|^2
	+ \cL_\text{others} \label{eq:setup}
\end{align}
where $\lambda$ is a coupling constant assumed to be perturbative, 
$\cL_\text{kin}$ denotes canonical kinetic terms
and $\cL_\text{others}$ represents the other light fields including gauge bosons.
Note that the scalar $\phi$ is assumed to interact with other light fields only through $\lambda^2 \phi^2 |\chi|^2$.
A complex scalar $\chi$ is assumed to be charged under some gauge group (\textit{e.g.}, SM gauge group)
and lighter than $\phi$ in the vacuum and its bare mass is neglected in the following.
We also assume that $\chi$ can decay into other light particles in the presence of non-zero
expectation value of $\phi$, for instance via a Yukawa interaction.
Through these gauge and Yukawa interactions, $\chi$ field has contact with the other light particles.
For simplicity, we consider the case where the gauge and Yukawa couplings, $g$ and $y$, are roughly 
the same orders $g \sim y$ and
the coupling between $\phi$ and $\chi$ is small $\lambda \ll \alpha := g^2 / (4\pi)$.
A typical example within this toy model is the singlet extension of the SM
where $\phi$ is a singlet and $\chi$ is SM Higgs boson, 
which is charged under SU(2)$_\text{W} \times$U(1)$_\text{Y}$ and  has the large top Yukawa coupling.

Throughout this paper, we consider the case where initially $\phi$ is displaced far from its potential minimum.
That is a typical situation for $m_\phi \ll H_\text{inf}$ with $H_\text{inf}$ being the Hubble parameter during
inflation. 
The evolution of scalar condensate is governed by the following effective equation of motion:
\begin{align}
	\ddot \phi + \left[ 3H (T) + \Gamma_\phi (\phi;T) \right]  \dot \phi+ \frac{\der V_\text{eff} (\phi;T)}{\der \phi} = 0,
	\label{eq:cgeq_z2}
\end{align}
where $T$ is the temperature of ambient plasma,
$H$ is the Hubble parameter, $V_\text{eff}$ is the effective potential that imprints 
finite density corrections and $\Gamma_\phi$ is the dissipation rate of oscillating scalar.
$\phi$ begins to oscillate when the Hubble parameter becomes comparable to its effective
mass that encodes the finite density correction: $m_\phi^{\text{eff}} := [\der V_\text{eff} / (\phi \der \phi)]^{1/2}$.
If the oscillating $\phi$ condensation is completely broken into $\phi$ particles
and thermalizes due to the dissipative term $\Gamma_\phi$
before the number changing annihilation process decouples from thermal equilibrium,
then the remnant of $\phi$ fields is determined by the standard thermal freeze-out
irrespective of its initial condition.
On the other hand, if the oscillating scalar survives, then the present abundance of scalar field
may depend on its initial condition.
Therefore, to predict the cosmological fate of oscillating scalar field with $Z_2$ symmetry,
we have to compare the time scales relevant to its relaxation with that of cosmic expansion.

In Sec.~\ref{sec:dyn_scalar}, we will not discuss the technical details to derive Eq.~\eqref{eq:cgeq_z2},
rather intuitively explain 
the basic results that will be derived in Sec.~\ref{sec:app} and concentrate on the scalar field dynamics.
In Sec.~\ref{sec:app},
we discuss the relation of the coarse-grained equations [including Eq.~\eqref{eq:cgeq_z2}] that we use in Sec.~\ref{sec:dyn_scalar}
with the Schwinger-Dyson (Kadanoff-Baym) equations on Closed Time Path (CTP) in detail.
In particular, we  evaluate the dissipation rate for oscillating scalar $\Gamma_\phi$  by using the CTP formalism.
Readers who are not interested in technical details can skip Sec.~\ref{sec:app}
and proceed to Sec.~\ref{sec:conc}. 
Sec.~\ref{sec:conc} is devoted to the conclusions and discussion.

\section{Dynamics of oscillating scalar field} 
\label{sec:dyn_scalar}

In this section, we do not attempt to derive the coarse-grained equations [including Eq.~\eqref{eq:cgeq_z2}]
and to perform the detailed computation of dissipation rate $\Gamma_\phi$.
Readers who are interested in these technical issues can find the detailed discussion in Sec.~\ref{sec:app}.
Instead, let us concentrate on the evolution of scalar condensate and its cosmological fate:
whether or not 
the coherently oscillating scalar can completely dissipate its energy against the cosmic expansion.

\subsection{Beginning of oscillation} 
\label{sec:}

In this section, let us briefly discuss the time when the scalar field starts to oscillate.
The following arguments closely follow Ref.~\cite{Moroi:2013tea}.
See also Sec.~\ref{sec:app}.

If there is a background plasma produced from the inflaton,
the beginning time may be affected since the effective potential is modified
due to the presence of background thermal plasma.
Since the $\chi$ field becomes massless at the origin of $\phi$'s potential,
there might be a sink.
Such effects are imprinted in the free energy at one-loop level as:
\begin{align}
	V_\text{1-loop} = N_\text{d.o.f.} \times \frac{T^4}{\pi^2}
	\int^\infty_0 dz\, z^2 \ln \left[ 1- e^{-\sqrt{z^2 + \lambda^2 \phi^2 / T^2}} \right]
\end{align}
where $N_\text{d.o.f.}$ is the number of $\chi$ particles normalized by one complex scalar.
For $\lambda |\phi| < T$, this term leads to the ``\textit{thermal mass}''~\cite{Dolan:1973qd}:
\begin{align}
	V_\text{1-loop} = N_\text{d.o.f.} \times \frac{ \lambda^2 T^2 }{12}\phi^2 + \cdots \label{eq:th_mass}
\end{align}
at the leading order in $\lambda$ [See also Eq.~\eqref{eq:th_mass_2}].

On the other hand, if $\lambda |\phi| \gg T$, this term rapidly vanishes due to the
Boltzmann suppression.
However, even in this case, the free energy depends on the field value $\phi$ 
via higher loop contributions.
Recalling that the free energy of hot plasma has a contribution
proportional to $g^2(T) T^4$ with $g(T)$ being the gauge coupling constant
at temperature $T$ and that the gauge coupling constant below the scale $\lambda |\phi|$ 
depends on $\phi$ logarithmically,
one finds the ``\textit{thermal logarithmic}'' potential~\cite{Anisimov:2000wx}:
\begin{align}
	V_\text{th-log} = a_\text{L} \alpha^2 (T) T^4 \ln \left[ \lambda^2 \phi^2/T^2 \right]
	\label{eq:th_log}
\end{align}
with $a_\text{L}$ being an order one constant.

Therefore, the thermal free energy that may affect 
the onset of $\phi$'s oscillation
can be parametrized as
\begin{align}
	V_\text{th} = 
	\begin{cases}
		a_\text{M} \lambda^2 T^2 \phi^2 & \text{for}~~\lambda |\phi| < T \\[5pt]
		a_\text{L} \alpha^2 (T) T^4 \ln \left[ \lambda^2 \phi^2 / T^2 \right]
		& \text{for}~~\lambda |\phi| > T
	\end{cases}
\end{align}
with $a_\text{M/L}$  being order one constants.
The effective potential for the scalar condensation is given by the sum: $V_\text{eff} := m_\phi^2 \phi^2/2 + V_\text{th}$.
The scalar field begins to oscillate when the Hubble parameter becomes
comparable to its effective mass:
\begin{align}
	H_\text{OS} \simeq \text{max} 
	\left[ m_\phi, \lambda T~(\text{for}~~\lambda |\phi| < T),
	\alpha T^2 / |\phi| ~(\text{for}~~\lambda |\phi| > T) \right]
\end{align}
where $H_\text{OS}$ is the Hubble parameter at the beginning of oscillation,
with the order one constants $a_\text{M/L}$ being neglected for simplicity.
It depends on the following parameters; the couplings $\lambda$ and $\alpha$, the initial amplitude $\phi_i$
and the reheating temperature $T_\text{R}$
when and by which term the scalar field begins to oscillate~\cite{Mukaida:2012qn,Moroi:2013tea}.

It is noticeable that if the oscillation of scalar field is so slow that
the $\chi$ particles can closely track the thermal equilibrium, then
the scalar field also oscillates with this thermal free energy
(See also Sec.~\ref{sec:cgeq_mf}).

Note that we have neglected the Coleman-Weinberg correction, which may give a dominant zero-temperature potential 
$V \propto \lambda^4 \phi^4$ at a large field value [see Eq.~(\ref{eq:th_mass_phi})].
It would modify the scalar dynamics in following ways.
First, the initial field value $\phi_i$ cannot be arbitrary large and it is bounded as $\phi_i \lesssim H_{\rm inf} / \lambda^2$
with $H_{\rm inf}$ being the Hubble parameter during inflation. 
Second, if it oscillates with $\phi^4$ potential, its energy density behaves as radiation rather than non-relativistic matter.
Third, there can be a non-perturbative self particle production effect due to $\phi^4$ interaction.
All these effects tend to reduce the scalar energy density compared with the case without $\phi^4$ potential.
Therefore, the scalar energy density estimated in the following should be regarded as a maximally possible one.
Actually the Coleman-Weinberg correction of $V \propto \lambda^4 \phi^4$ does not appear if the theory is embedded in supersymmetry.
We will discuss that, even in such a case, the scalar energy density can be completely dissipated.

\subsection{Non-thermal/Thermal dissipation} 
\label{sec:diss}

In this section, we analyze how the oscillating scalar field
dissipates its energy into background plasma intuitively.
See Sec.~\ref{sec:app} for details.
There are roughly two classes of dissipation.
(i) The oscillating scalar field loses its energy via the non-perturbative production~\cite{Kofman:1994rk} (See also Sec.~\ref{sec:np_prod}).
(ii) The oscillating scalar field loses its energy via the thermal dissipation due to
the abundant background thermal plasma~\cite{Berera:1995ie,Yokoyama:2004pf,Drewes:2010pf} (See also Sec.~\ref{sec:cgeq_mf}).
Let us discuss them in the following.

\subsubsection{Non-thermal dissipation} 
\label{sec:np}

After the onset of oscillation, the scalar field oscillates around its
effective potential minimum, and hence
the coupled $\chi$ particles have a time dependent dispersion relation:
$\omega_\chi^2 = \bm{p}^2 + m_{\chi,\text{th}}^2 + \lambda^2 \phi^2 (t)$.\footnote{
	Here we assumed that the background plasma can remain in thermal equilibrium.
	Otherwise, the screening mass of $\chi$ particles may not be described by
	the ``thermal mass''.
}
The non-perturbative particle production occurs when the adiabaticity of $\chi$ particles is
broken down $|\dot \omega_\chi / \omega_\chi^2| \gg 1$ and
the $\phi$'s amplitude $\tilde \phi$ is so large that 
$\lambda \tilde \phi \gg m_\phi$~\cite{Mukaida:2012qn,Mukaida:2012bz}:\footnote{
	One can show that if the scalar oscillates with the thermal free energy,
	the non-perturbative production does not occur for $\lambda \ll \alpha$~\cite{Mukaida:2012qn}.
	See also Fig.~\ref{fig:gamma} in the Appendix.
}
\begin{align}
	\lambda \tilde \phi \gg \text{max} \left[ m_\phi, \frac{m_{\chi,\text{th}}^2}{m_\phi} \right].
	\label{eq:np_cond}
\end{align}
Note that the non-perturbative production is ``blocked'' if the temperature of background
plasma is so high that $\lambda \tilde \phi m_\phi \lesssim m_{\chi,\text{th}}^2$.

Throughout this paper, the $\chi$ particle is assumed to decay into other light particles
with a fairly large rate.
Hence, the non-perturbatively produced $\chi$ particles tend to decay completely
well before the oscillating scalar moves back to its origin again~\cite{Felder:1998vq}.
This is the case for
\begin{align}
	y^2 \lambda \tilde \phi \gg m_\phi
	\label{eq:decay_cond}
\end{align}
with the decay rate of $\chi$ being $\Gamma^\chi_\text{dec} \sim y^2 \lambda |\phi (t)|$,
and we mainly concentrate on this case in the following.\footnote{
	Otherwise, the parametric resonance occurs due to the induced emission effect from
	the previously produced $\chi$ particles. See also the discussion in Appendix.~\ref{sec:narrow}.
}
If Eq.~\eqref{eq:decay_cond} is satisfied, the scalar field loses its energy via the perturbative decay of
$\chi$ into other light particles for each crossings of $|\phi| < \phi_\text{NP}$.
It is noticeable that the decay is dominated at the outside of the non-adiabatic region
and hence the usage of ``particle decay'' is justified a posteriori.
The effective dissipation rate can be evaluated as
\begin{align}
	\Gamma_\phi^\text{NP} \sim N_\text{d.o.f} \times
	 \frac{\lambda^2 m_\phi}{2 \pi^4 |y|}. 
	 \label{eq:diss_np}
\end{align}

The produced light particles via the decay of ``heavy'' $\chi$ typically have large momenta compared
to the thermal distribution. This is the so-called under occupied situation and its thermalization is
extensively studied (See Ref.~\cite{Kurkela:2011ti} for instance).
If the thermalization time scale of these light particles is much faster than the oscillation period of scalar field,
one can easily track the evolution of oscillating scalar/plasma system with 
assuming that the background plasma remains in thermal equilibrium~\cite{Mukaida:2012bz}.

\subsubsection{Thermal dissipation} 
\label{sec:th_diss}

When the condition Eq.~\eqref{eq:np_cond} is violated (or the outside of non-adiabatic region),
the particle concept of $\chi$ field is well defined in the WKB sense.
In this regime, the oscillating scalar dissipates its energy due to the presence of 
abundant background plasma.

It is practically difficult to follow the evolution of oscillating scalar/plasma system in a general setup.
However, there are particular cases where the equations become rather simple as shown in Sec.~\ref{sec:cgeq_mf}.
(a) The oscillation of scalar field is so slow that $\chi$ can be assumed to be in thermal equilibrium at each field value.
(b) The amplitude of oscillating scalar is smaller than thermal mass of $\chi$ and hence $\chi$ remains in thermal equilibrium.
For clarity, we have discussed the relation of the coarse-grained eqs.\ with Schwinger-Dyson eqs.\ 
on CTP and summarized the technical details in Sec.~\ref{sec:app}. (See also Ref.~\cite{Mukaida:2012qn}.)
In this section, we will not repeat the technical details but summarize basic results to study the coarse-grained dynamics of
oscillating scalar, and then see how the oscillating scalar/plasma system evolves.
\\

\noindent \textbf{The Case (a) $\bm{m_\phi \ll \alpha T}$}\textbf{:}

Before going into details, let us recall that since we will study the case where the scalar field oscillates slowly in the following,
the naive ``perturbative decay'' of scalar field into quasi-particles in thermal plasma is not possible due to the
relatively large screening mass of would-be decay products compared with the effective mass of scalar field.
However, even in this case, the oscillating scalar can dissipate its energy into thermal background;
roughly speaking, via multiple scatterings.

First, let us consider the case of $\lambda |\phi (t)| \gg T$.
Since the $\chi$ particles are assumed to decay into other light particles in thermal plasma
immediately, there are no $\chi$ particles in this case.
Nevertheless, the scalar field interacts with background thermal plasma via a dimension 5 operator
that can be obtained from integrating out heavy $\chi$ fields,
and through this interaction the scalar field dissipates its energy.
The dimension five operator is given by $(\delta \phi/\bar\phi) F_{a\mu\nu}F^{a\mu\nu}$
where $\bar \phi = \phi (\bar t)$ and $\delta \phi (t; \bar t) = \phi (t) - \bar\phi$,
and this term induces the following dissipation factor~\cite{Anisimov:2000wx,Bodeker:2006ij,Moroi:2012vu,Mukaida:2012qn}:
\begin{align}
	\Gamma^\text{(dim5)}_\phi \sim \frac{b \alpha^2 T^3}{\phi^2 (\bar t)},
	\label{eq:dim5}
\end{align}
where
\begin{align}
	b = \left(  \frac{\text{T} (r)}{16 \pi^2}  \right)^2 \frac{\( 12 \pi \)^2}{\ln \alpha^{-1}};
\end{align}
typically, $b \sim 10^{-3}$. $\text{T} (r)$ is the normalization of representation $r$:
$\text{T}(r) \delta^{ab} = \Tr [t^a (r) t^b (r)]$.

Next, let us consider the case of $\lambda |\phi (t)| < T$.
In this regime, the $\chi$ particles can be approximated to be as abundant as the thermal distribution since
the scalar field is assumed to oscillate slow enough.
Actually, this is the case for $\lambda \tilde \phi m_\phi \ll m_{\chi,\text{th}}^2$ [See Eq.~\eqref{eq:np_cond}].
The time scale $\delta t$, during which the oscillating scalar passes through the region $\lambda |\phi (t)| < T$,
can be estimated as $\delta t \sim T/ (\lambda \tilde \phi m_{\phi})$.
Hence, the $\chi$ particles can be thermally populated since 
$\Gamma_\text{prod} \delta t \sim y^2 T^2 / (\lambda \tilde \phi m_\phi) \gg 1$.
There are two processes of energy transportation from oscillating scalar
into light particles, so let us discuss them in turn.

The first process is a counterpart of decay at vacuum via the effective three-point interaction ($\lambda^2\langle\phi\rangle\phi \chi^2$),
which is proportional to $\lambda^4 \phi^2/m_\phi$.
In the thermal background plasma, such a perturbative decay is not possible due to the large screening mass of $\chi$.
Instead, this dissipation rate is modified to Eq.~\eqref{eq:diss_ann}:
\begin{align}
	\Gamma^{(a)}_\phi \sim N_\text{d.o.f.} \times \frac{\lambda^4 \phi^2 (\bar t)}{ \pi^2 \alpha T}.
	\label{eq:diss_ann_2}
\end{align}
Here we have approximated the integrand factor
to estimate the dissipation rate for $\lambda |\phi| < T$. On the other hand,
in the case of $\lambda |\phi| > T$, the exponential factor in the integrand dominates
and this dissipation rate is exponentially suppressed.

The second one is a scattering where the oscillating scalar condensation is scattered off by abundant $\chi$ particles,
namely $\phi \chi \to \varphi\chi$ with $\varphi$ being a $\phi$-particle ($\phi$ is the oscillating scalar condensation).
The dissipation rate for this process can be estimated as Eq.~\eqref{eq:diss_scat_2}:
\begin{align}
	\Gamma_\phi^{(b)} \sim N_\text{d.o.f.} \times \frac{\lambda^4 T^3}{12 \pi  m_{\chi,\text{th}}^2}.
	\label{eq:diss_scat_2_2}
\end{align}

Importantly, this process alone cannot heat the background plasma, rather
it drains energy from the background plasma and produces $\varphi$ particles instead.
However,
as we will see later, whenever this scattering process [Eqs.~\eqref{eq:diss_scat_2_2}]
dominates the dissipation of oscillating scalar,
the $\chi$ particles are relativistic and the energy of oscillating scalar is smaller than that of relativistic particles $T^4$
for $m_\phi < T$.
Hence, the oscillating scalar is expected to dissipate its energy without cooling the background plasma
and the produced $\varphi$ particles soon participate in the thermal plasma via interactions imprinted in Eq.~\eqref{eq:boltz_4pt}.
See  Sec.~\ref{sec:dyn} for detail.
\\

\noindent \textbf{The Case (b) $\bm{\lambda \tilde \phi \ll m_{\chi,\text{th}}}$}\textbf{:}

In this case,
in contrast to the case (a), the mass of oscillating scalar can be larger than $\alpha T$
while the background plasma including $\chi$ particles is expected to remain in thermal equilibrium
for the following reasons.
First, the field value dependence of $\chi$'s mass can be neglected since $\lambda \tilde \phi \ll m_{\chi,\text{th}}$.
Second, the energy transportation rate from the oscillating scalar to the background plasma
is much smaller than the typical interaction rate of thermal plasma.
Finally, both the broad and narrow resonances are not likely to occur in our case.
See the discussion at the beginning of Sec.~\ref{sec:cgeq_mf_small}.

In the case of $m_\phi \ll \alpha T$, the dissipation rates are the same as the case (a),
and hence let us concentrate on the case: $\alpha T \ll m_\phi \ll T$.
Similar to the case (a), there are two processes of energy transportation.

The first one is a counterpart of decay at vacuum via the effective three-point interaction.
If the mass of oscillating scalar is smaller than the thermal mass of $\chi$,
the dissipation rate is the same as the case (a) and it is given by Eq.~\eqref{eq:diss_ann_2}.
On the other hand, if the oscillating scalar is heavier than the $\chi$ quasi-particles,
then a perturbative decay (annihilation) is kinematically allowed.
Hence, the dissipation rate is given by Eq.~\eqref{eq:diss_ann_s}:
\begin{align}
	\Gamma\a_\phi =
	N_\text{d.o.f.} \times \frac{\lambda^4 \phi^2}{8 \pi m_\phi} \sqrt{1 - \frac{m_{\chi,\text{th}}^2}{m_\phi^2}} 
	\left[ 1 + 2 f_\text{B} (m_\phi) \right] \theta \( m_\phi^2 - m_{\chi,\text{th}}^2 \).
\end{align}

The second one is the scattering by the abundant $\chi$ particles.
In the former case (a), the final particles $\varphi \chi$ are almost collinear in the
rest frame of thermal plasma, and hence the phase space is suppressed.
In contrast, for $\alpha T \ll m_ \phi \ll T$,
the dissipation rate is given by Eq.~\eqref{eq:diss_scat_s}
\begin{align}	
	\Gamma\b_\phi 
	\sim N_\text{d.o.f.}\times \frac{\lambda^4 T^2}{48 \pi m_\phi}
\end{align}
where there are no $\varphi$ particles.
If the $\varphi$ particles are as abundant as thermal one,
then the rate increases by a factor of $3/2$.

\subsection{Dynamics of oscillating scalar field} 
\label{sec:dyn}

Then, using the obtained equations, let us study the dynamics of oscillating scalar with $Z_2$ symmetry.
The coarse-grained equation is given by
\begin{align}
	\ddot \phi + \left[ 3H (T) + \Gamma_\phi (\phi;T) \right]  \dot \phi+ \frac{\der V_\text{eff} (\phi;T)}{\der \phi} = 0
\end{align}
where $H$ is the Hubble parameter and
\begin{align}
	V_\text{eff}(\phi;T) = \frac{1}{2}m_\phi^2 \phi^2 + V_\text{th}(\phi;T).
\end{align}
Since we are interested in the evolution of energy density,
it is convenient to consider quantities averaged over a time interval
that is longer than the oscillation period but shorter than the Hubble and dissipation rate.

Until the averaged-dissipation rate, $\Gamma_\phi^\text{eff} \equiv \overline{\Gamma_\phi \dot \phi^2}/\overline{\dot \phi^2}$
with $\overline{\cdots}$ being the oscillation time average,
becomes as large as the Hubble parameter,
the oscillating scalar mainly loses its energy because of the Hubble expansion.
In that regime, we can obtain the following scaling solutions for the amplitude of oscillating scalar~\cite{Mukaida:2012qn}:
\begin{align}
	\tilde \phi \propto
	\begin{cases}
		a^{-3/2} &\text{for zero temperature mass},\\
		a^{-3/2} T^{-1/2} &\text{for thermal mass},\\
		a^{-3} T^{-2} &\text{for thermal log},
	\end{cases}
\end{align}
with $a$ being the scale factor.

The oscillating scalar condensation is expected to evaporate when the averaged-dissipation rate becomes comparable to
the Hubble parameter: $\Gamma^\text{eff}_\phi \simeq H$.
To estimate the evaporation time, we have to know the averaged-dissipation factor in the various regimes.
Hence, let us study the averaged-dissipation rate in the following.
\begin{description}
	\item[$\bullet$ The thermal mass:]
		In this case, the oscillating scalar dissipates its energy via the effective three point interaction [Eq.~\eqref{eq:diss_ann_2}]
		and the scatterings: $\phi \chi \to \varphi \chi$ [Eq.~\eqref{eq:diss_scat_2_2}].
		Taking the time-average, one finds the dissipation factor as
		\begin{align}
			\Gamma_\phi^\text{eff} \sim N_\text{d.o.f.} \times 
			\begin{cases}
				\cfrac{\lambda^4 \tilde \phi^2}{\alpha T} & \text{for}~~ T < \tilde \phi \(\ll \frac{T}{\lambda}\),  \\[10pt]
				\cfrac{\lambda^4 T}{\alpha} & \text{for}~~ \tilde \phi < T.
			\end{cases}
			\label{eq:av_diss_thmass}
		\end{align}
		Here we have dropped numerical factors for brevity and approximated the thermal mass of 
		$\chi$ as $m_{\chi,\text{th}} \sim gT$.
		Note that this dissipation rate is always smaller than the thermal mass $m_{\phi,\text{th}}\sim \lambda T$ since
		we have $\lambda \ll \alpha$.
	\item[$\bullet$ The thermal log:]
		In this case,
		for the large field value regime ($\lambda |\phi (t)| \gg T$), the dissipation is caused by scatterings with gauge bosons in thermal plasma via
		the dimension five parameter [Eq.~\eqref{eq:dim5}]. On the other hand,
	 	for the small field value regime ($\lambda |\phi (t)| < T$), it is caused by the effective three point interaction [Eq.~\eqref{eq:diss_ann_2}]
		and the scatterings: $\phi \chi \to \varphi \chi$ [Eq.~\eqref{eq:diss_scat_2_2}].
		By taking the time-average, one finds that the dissipation is dominated by the effective three point interaction
		and the averaged-dissipation rate is given by\footnote{
			Note that the dissipation is computed in two limits $\lambda |\phi| \gg T$  and $\lambda |\phi| \ll T$,
			and hence we have some ambiguities in the intermediate regime.
		}
		\begin{align}
			\Gamma_\phi^\text{eff} \sim N_\text{d.o.f.} \times \frac{\lambda T^2}{\tilde \phi}.
			\label{eq:av_diss_thlog}
		\end{align}
		Here we have dropped numerical factors for brevity and approximated the thermal mass of $\chi$ as 
		$m_{\chi,\text{th}} \sim gT$.
		Again, note that this dissipation rate is smaller than the effective mass term
		$\alpha T^2/\tilde \phi \gg \Gamma^\text{eff}_\phi$
		since we have $\lambda \ll \alpha$.
	\item[$\bullet$ The zero temperature mass:]
		In most cases,\footnote{
		At the very time 
		when the non-perturbative production terminates,
		the following condition is satisfied: $gT < k_\ast < T$.
		Since this implies $\phi_\text{NP} < T/\lambda$,
		the thermal dissipation rate Eq.~\eqref{eq:diss_ann_2} may become comparable to the non-perturbative production rate
		at this short transition time interval with $gT < k_\ast < T$.
		However, in ``most cases'' $T < k_\ast$, the non-perturbative particle production dominates the averaged-dissipation
		rate.
		See Appendix.~B of Ref.~\cite{Mukaida:2012bz}.
		} if the non-perturbative particle production occurs, then the dissipation of oscillating scalar is
		dominated by this process. Thus, in the case of 
		$\lambda \tilde \phi m_\phi \gg \max[m_{\chi,\text{th}}^2,m_\phi^2]$, 
		the averaged-dissipation rate is given by
		\begin{align}
			\Gamma_\phi^\text{eff} \sim N_\text{d.o.f.} \times \frac{\lambda^2 m_\phi}{ 2 \pi^4 |y|}.
		\end{align}

		Then, we concentrate on the case of $\lambda \tilde \phi m_\phi \lesssim \max[m_{\chi,\text{th}}^2,m_\phi^2]$ where
		 the non-perturbative production is blocked.
		First, let us consider the case of $m_\phi \ll \alpha T$.
		In the same way as the thermal log potential, 
		for the large field value regime ($\lambda |\phi (t)| \gg T$), the dissipation is caused by 
		the dimension five parameter [Eq.~\eqref{eq:dim5}], and for the small field value regime ($\lambda |\phi (t)| < T$), 
		it is caused by the effective three point interaction [Eq.~\eqref{eq:diss_ann_2}]
		and the scatterings [Eq.~\eqref{eq:diss_scat_2_2}].
		Hence, the averaged-dissipation factor with $m_\phi \ll \alpha T$ is given by
		\begin{align}
			\Gamma_\phi^\text{eff} \sim N_\text{d.o.f.} \times
			\begin{cases}
				\cfrac{\lambda T^2}{\alpha \tilde \phi} & \text{for}~~\frac{T}{\lambda} \ll \tilde \phi, \\[15pt]
				\cfrac{\lambda^4 \tilde \phi^2}{\alpha T} & \text{for}~~T < \tilde \phi \ll \frac{T}{\lambda}, \\[15pt]
				\cfrac{\lambda^4 T}{\alpha} & \text{for}~~\tilde \phi < T,
			\end{cases}
			\label{eq:av_diss_small_m}
		\end{align}
		where we have dropped numerical factors.
		
		Second, we consider the case: $\alpha T \ll m_\phi \ll T$.
		This implies $\lambda \tilde \phi < T$ since we consider 
		$\lambda \tilde \phi m_\phi \lesssim \max[m^2_{\chi,\text{th}},m_\phi^2]$.
		Particularly, let us concentrate on a parameter region $\lambda \tilde \phi \ll m_{\chi,\text{th}} \sim gT$
		because one can assume that the background plasma including $\chi$ is kept in thermal equilibrium.
		See Sec.~\ref{sec:cgeq_mf_small} for details.
		In the case of $\lambda \tilde \phi \ll gT$ and $\alpha T \ll m_\phi \ll T$, the averaged-dissipation rate can be
		expressed as [See Eq.~\eqref{eq:diss_scat_2_2}]:
		\begin{align}
			\Gamma_\phi^\text{eff}& \sim N_\text{d.o.f.} \times
			\begin{cases} 
				\cfrac{\lambda^4 \tilde \phi^2}{\alpha T} &\text{for}~~
				\sqrt{ \frac{\alpha T}{m_\phi} }\, T < \tilde \phi \ll  \frac{gT}{\lambda}\\[15pt]
				\cfrac{\lambda^4 T^2}{m_\phi} &\text{for}~~ \tilde \phi < \sqrt{ \frac{\alpha T}{m_\phi} }\, T
			\end{cases}&\text{with}~~\alpha T \ll m_\phi \ll gT; \label{eq:av_diss_mid_m}
		\end{align}
		\begin{align}
			\Gamma_\phi^\text{eff}& \sim N_\text{d.o.f.} \times
			\begin{cases} 
				\cfrac{\lambda^4 \tilde \phi^2}{m_\phi} &\text{for}~~T < \tilde \phi \ll  \frac{gT}{\lambda}\\[15pt]
				\cfrac{\lambda^4 T^2}{m_\phi} &\text{for}~~ \tilde \phi < T
			\end{cases}&\text{with}~~gT \ll m_\phi \ll T. \label{eq:av_diss_large_m}
		\end{align}
		Here we have dropped numerical factors for brevity.
\end{description}

Several remarks are in order:
\begin{itemize}
	\item First, in any case, the amplitude dependent dissipation rate that is proportional to $\tilde \phi^2$ alone cannot
	fully transport the energy of oscillating scalar into the background plasma 
	since the dissipation rate decreases more rapidly than the Hubble parameter.
	The existence of scatterings in the small amplitude regime ($\tilde \phi < T$) is essential for the oscillating scalar
	to transport its energy completely.
	\item Second, as mentioned in the pervious Sec.~\ref{sec:th_diss},
	the scatterings $\phi \chi \to \varphi \chi$ alone cannot heat the background plasma, rather
	it drains energy from thermal plasma and produces $\varphi$ particles instead.
	However, as one can see from Eqs.~\eqref{eq:av_diss_thmass}, \eqref{eq:av_diss_small_m}, \eqref{eq:av_diss_mid_m}
	and \eqref{eq:av_diss_large_m},
	the scatterings dominate the dissipation rate for $\tilde \phi < T$ at least.
	At that regime, the $\chi$ particles are relativistic and 
	the energy fraction of oscillating scalar for $m_\phi < T$ is smaller than that of relativistic particles $T^4$.
	Thus, we expect that the oscillating scalar dissipates its energy
	without cooling the thermal plasma.
	\item Third, Eq.~\eqref{eq:av_diss_large_m} implies the critical value of coupling $\lambda$ that detemines
	whether or not the oscillating scalar can dissipate its energy completely.
	This is because, the dissipation rate that is proportional to $T^2$ in 
	Eqs.~\eqref{eq:av_diss_mid_m} and \eqref{eq:av_diss_large_m} 
	cannot exceed the Hubble rate that is also proportional to
	$T^2$ in the radiation-dominated era.\footnote{
		In this case, $\tilde \phi < T$ and $m_\phi < T$, the energy density of oscillating scalar
		is at most comparable to radiation.
	}
	Hence, in order for the oscillating scalar to successfully dissipate its energy,
	it should evaporate before the temperature decreases as low as $\alpha T \lesssim m_\phi$.
	This implies the following critical value:
	\begin{align}
		\lambda_c \sim \left[ \frac{m_\phi}{M_\text{pl}} \right]^{1/4}.  \label{lambda_c}
	\end{align}
	Below this critical value, the oscillating scalar survives from the thermal dissipation.
	\item
	Fourth, even after the coherently oscillating scalar disappears,
	the phase space distribution of produced $\varphi$ particles is still shapely 
	dominated by the IR momentum that is much smaller than $T$.
	Let us estimate the time scale $\delta t_\text{UV}$ which the $\varphi$'s distribution takes to evolve towards UV-regime $T$
	via the interactions imprinted in Eq.~\eqref{eq:boltz_4pt}.
	First, the typical gain of the momentum of $\varphi$ in each scatter $\varphi \chi \to \varphi \chi$, denoted by $\delta p$, 
	is given by $\delta p \sim E_\varphi / g^2$ as long as $E_\varphi \lesssim gT$.
	Thus the typical energy of $\varphi$ just after the dissipation is given by $\sim m_\phi / g^2$.
	Next, the typical scattering rate for $m_\phi < \alpha T$ is $\Gamma_\text{scat} \sim \lambda^4 T /\alpha$.
	Thus the momentum of $\varphi$ grows in a time scale of $\delta t_\text{UV} \sim g^2 / \Gamma_\text{scat}$.
	It is comparable to the Hubble time scale at the completion of dissipation.
	Therefore, whenever the oscillating scalar can completely dissipate its energy,
	the produced $\varphi$ particles soon participate in thermal plasma.
	\item Finally, in the case of $m_\phi > T$ with $\lambda \tilde \phi < m_\phi$, the annihilation of $\phi$ is thermally decoupled
	before the coherently oscillating scalar is broken into relativistic particles.
	Thus, the coherently oscillating scalar survives and tends to dominate the Universe.
\end{itemize}

\subsection{Numerical result} 
\label{sec:num}

\begin{figure}[htbp]
\begin{center}
\includegraphics[scale=1.0]{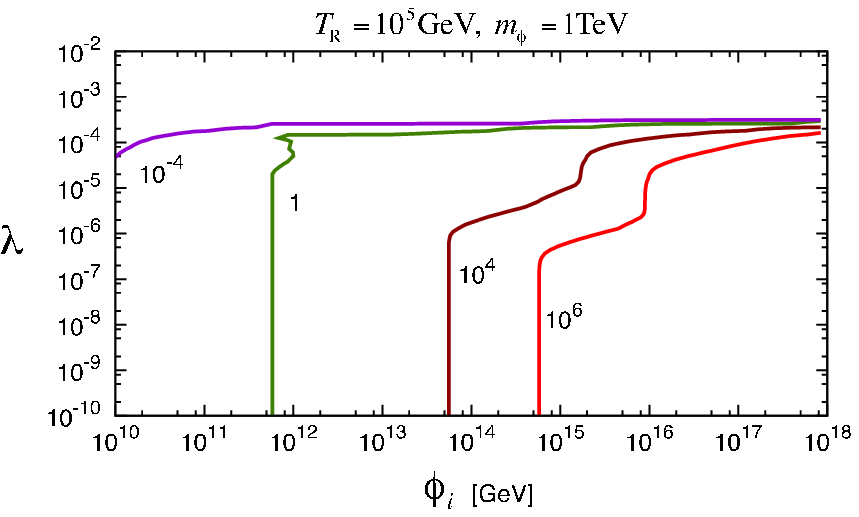}
\vskip 1cm
\includegraphics[scale=1.0]{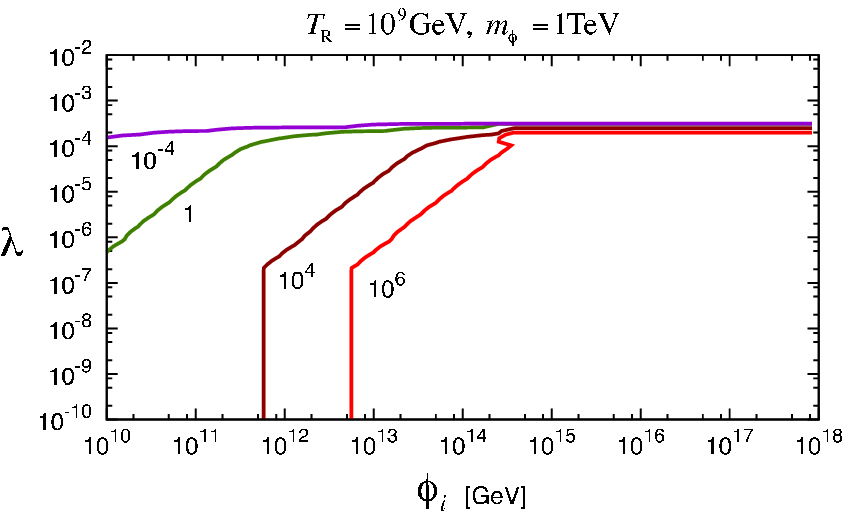}
\vskip 1cm
\includegraphics[scale=1.0]{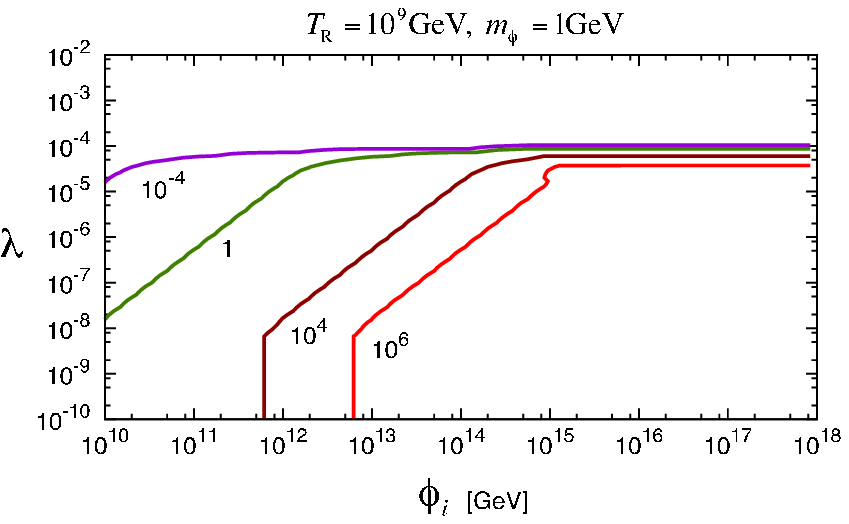}
\caption{ 
	Contours of the abundance of the coherently oscillating scalar field in units of DM abundance
	on $(\phi_i, \lambda)$ plane. We have taken $(m_\phi, T_{\rm R}) = (1\,{\rm TeV}, 10^5\,{\rm GeV})$ (top),
	$(m_\phi, T_{\rm R}) = (1\,{\rm TeV}, 10^9\,{\rm GeV})$ (middle)
	and $(m_\phi, T_{\rm R}) = (1\,{\rm GeV}, 10^{9}\,{\rm GeV})$ (bottom).
}
\label{fig:cont}
\end{center}
\end{figure}

Now we are in a position to calculate the scalar dynamics including all the effects mentioned before.
The results of numerical calculation are shown in Fig.~\ref{fig:cont}, where we have plotted
contours of the abundance of the coherently oscillating scalar field at present in units of DM abundance on $(\phi_i, \lambda)$ plane: 
$\rho_\phi / \rho_{\rm DM}$ with $\rho_{\rm DM}$ representing the DM abundance.
We have taken $(m_\phi, T_{\rm R}) = (1\,{\rm TeV}, 10^5\,{\rm GeV})$ (top),
$(m_\phi, T_{\rm R}) = (1\,{\rm TeV}, 10^9\,{\rm GeV})$ (middle)
and $(m_\phi, T_{\rm R}) = (1\,{\rm GeV}, 10^{9}\,{\rm GeV})$ (bottom).
The regions with $\rho_\phi / \rho_{\rm DM} < 1$ are allowed.
It is seen that, unless the initial amplitude is very small, 
the dissipation is efficient so that the scalar field energy density is efficiently dissipated into the radiation
if the coupling $\lambda$ is larger than the critical value (\ref{lambda_c}).

In this figure, we have only taken into account the coherent oscillation.
However, if the dissipation effect is strong enough, the scalar field is expected to be thermalized in the plasma with temperature
higher than the scalar mass.
After the temperature drops to $T\sim m_\phi/20$, the scalar particles decouple from thermal bath
and the resulting relic abundance is determined by its self annihilation cross section [See Eq.~\eqref{eq:boltz_4pt}]:
\begin{equation}
	\langle \sigma v\rangle_{\varphi\varphi \to \chi\chi} \simeq N_\text{d.o.f.} \times\frac{\lambda^4}{8\pi m_\phi^2}\sqrt{1-\frac{m_\chi^2}{m_\phi^2} }.
\end{equation}
The the relic abundance is estimated as
\begin{equation}
	\Omega_\varphi h^2 \simeq 8\times 10^{-1} 
	\left( \fr{1}{N_\text{d.o.f.}} \right) \left( \frac{0.1}{\lambda} \right)^4 \left( \frac{m_\phi}{100\,{\rm GeV}} \right)^2,
\end{equation}
where $\rho_{\rm cr}$ is the critical energy density at present.
Therefore, in order for the thermal relic abundance not to exceed the DM abundance, the coupling $\lambda$ must be fairly large. 
In such a case, the coherent oscillation is expected to be efficiently dissipated so that its contribution to the relic energy density
is safely neglected.

\section{Formalism} 
\label{sec:app}

Note that readers who are not interested in technical details can skip this section.

In this section, let us clarify the relation of the coarse-grained equations that
we use throughout this paper with the Schwinger-Dyson (Kadanoff-Baym) eqs.\ on CTP~\cite{Kadanoff:1962,Baym:1961zz}
derived from 2PI (two-particle irreducible) effective action~\cite{Cornwall:1974vz}:
the self consistent set of evolution equations for the mean field and two point correlators.
See~\cite{Chou:1984es,Berges:2004yj,calzetta2008nonequilibrium}  for reviews.
Though the obtained coarse-grained equations are formally equivalent to 1PI (one-particle irreducible)
open system computaion
with integrating out $\chi$ fields,
we believe that the following arguments may clarify approximations and their limitation
from the perspective of full evolution equations.

In the following, we denote the CTP contour $\cal{C}$~\cite{Schwinger:1960qe} 
ordered propagator as
\begin{align}
	G (x,y) = \left< T_{\mathcal{C}}\, \hat \phi (x) \hat \phi (y) \right>
	= \theta_{\cal C} (x^0,y^0) G_> (x,y) + \theta_{\cal C} (y^0,x ^0) G_< (x,y),
\end{align}
and the Jordan (spectral) and Hadamard (statistical) propagators as
\begin{align}
	G_{J/H} (x,y) = \left< \left[ \hat \phi (x), \hat \phi (y) \right]_{\mp} \right>
\end{align}
with $[\bullet, \bullet]_\mp$ being the commutator and anti-commutator respectively.\footnote{
	Here and hereafter we consider bosonic fields.
	In the case of fermionic fields, we have to take care of Grassmann nature of these fields.
}
This implies the following relation:
\begin{align}
	G(x,y) = \frac{1}{2} \left[ G_H (x,y) + \text{sgn}_{\cal C} (x^0,y^0) G_J (x,y) \right]
\end{align}
with $\mathrm{sgn}_{\cal C}$ being a sign function defined on the contour ${\cal C}$.
In the case of a spatially homogeneous system, the propagator depends on the difference of
two distinct spatial points, and hence it is convenient to perform the Fourier transformation:
\begin{align}
	G_\bullet (x^0,y^0; \bm{p}) 
	= \int d^3 (x-y)\, e^{- i \bm{p} \cdot ( \bm{x} - \bm{y})} G_\bullet (x,y).
\end{align}

If the propagator is given by the thermal one,
then one can further Fourier transform the Green function:
\begin{align}
	G_\bullet^\text{th} (P) = \int d t\, e^{i p_0 t} G_\bullet^\text{th} (t;\bm{p})
\end{align}
where $(P) = (p_0,\bm{p})$.
For the thermal propagators, we have the Kubo-Martin-Schwinger (KMS) relation~\cite{Kubo:1957mj}:
\begin{align}
	G_>^\text{th} (P) = e^{p_0/T} G_<^\text{th} (P).
\end{align}
This implies the following useful relations:
\begin{align}
	G_>^\text{th} (P)  & = \left[ 1 + f_\text{B} (p_0) \right] \rho_\text{th} (P), \label{eq:rel_kms_1}\\
	G_<^\text{th} (P) &= f_\text{B} (p_0)\rho_\text{th} (P), \label{eq:rel_kms_2}\\
	G_H^\text{th} (P) &= \left[ 1 + 2f_\text{B} (p_0) \right] \rho_\text{th} (P), \label{eq:rel_kms_3}
\end{align}
with the spectral density being $\rho_\text{th} (P) = G_J^\text{th} (P)$ and the Bose-Einstein distribution being $f_\text{B}$.

\subsection{Schwinger-Dyson (Kadanoff-Baym) eqs.} 
\label{sec:2pi}

To truncate the Schwinger-Dyson hierarchy systematically,
it is convenient to make use of 2PI effective action, which is defined as the double Legendre
transformation of one point and two point external sources [$J(x)$ and $J_2(x,y)$] that couple to
the fields in consideration as $J \cdot \Phi$ and $\Phi \cdot J_2 \cdot \Phi$~\cite{Cornwall:1974vz}.
In the following, let us consider the case where other light particles than $\phi$ and $\chi$
can remain in thermal equilibrium.
The applicability of this approximation is discussed in each cases later.
Hence, we will not write down the contribution from other light fields explicitly
unless otherwise stated, and
closely follow the discussion given in Ref.~\cite{Aarts:2007ye}.

The 2PI effective action of Eq.~\eqref{eq:setup} is given by~\cite{Aarts:2007ye}
\begin{align}
	\Gamma_\text{2PI}[\phi, G_\varphi, G_\chi]
	= S[\phi] &+ \frac{i}{2} \Tr\, G^{-1}_{\varphi,0} \cdot G_\varphi - \frac{i}{2} \Tr \ln G_\varphi
	- \frac{i}{2}	 \Tr\, G^{-1}_\varphi \cdot G_\varphi \nn[5pt]
	&+ i \Tr\, G^{-1}_{\chi,0} \cdot G_\chi - i \Tr \ln G_\chi
	- i \Tr\, G^{-1}_\chi \cdot G_\chi \nn[5pt]
	& + \Gamma_2 [\phi, G_\varphi, G_\chi]
\end{align}
where $S [\phi]$ represents the tree level action and the free propagators are defined as follows:
\begin{align}
	i G_{\varphi,0}^{-1}(x,y) &=  - \left[ \Box_x + m_\phi^2 \right] \delta_{\cal C} (x,y)\\[5pt]
	i G_{\chi,0}^{-1}(x,y) &=  - \left[ \Box_x + \lambda^2 \phi^2 (x) \right] \delta_{\cal C} (x,y),
\end{align}
and $\Gamma_2$ contains all the two-particle irreducible vacuum bubbles,
that depend on $\phi$ and $G_{\phi/\chi}$.
Here and hereafter possible gauge indices are suppressed for brevity unless otherwise stated.
By performing a weak coupling expansion, we can truncate the $\Gamma_2$ systematically\footnote{
	Here we have ignored three loop diagrams of the higher order in the coupling $\lambda$.
	This implies that $\lambda^2 \phi$ expansion should be controlled.
	Since the effective mass $\lambda^2 \phi^2$ is completely resummed,
	the $\chi$ becomes heavy at large field value of $\phi$. 
	Hence, their number density is quite suppressed
	in our case because the $\chi$ has thermal contact with other light particles and
	can decay immediately.
	Thus, the contribution from a large $\phi$ is expected to be suppressed for $\lambda |\phi |> T$.
}
\begin{align}
	i \Gamma_2 [\phi,G_\varphi,G_\chi] 
	=& -i \lambda^2 \int_{\cal C} d^4x\, G_\varphi (x,x) G_\chi (x,x)  \label{eq:local} \\
	&+ \frac{1}{2} (- 2 i \lambda^2)^2 \int_{\cal C} d^4x d^4 y\, 
	\phi (x) G_\varphi(x,y) G_\chi (x,y) G_\chi(y,x) \phi (y)  \label{eq:sunset} \\
	&+ 2 \frac{1}{2} (-i \lambda^2)^2 \int_{\cal C} d^4x d^4y\,
	G_\varphi (x,y)^2 G_\chi(x,y) G_\chi(y,x)  \label{eq:bball} \\
	&+\cdots. \nonumber
\end{align}

The Schwinger-Dyson (Kadanoff-Baym) eqs. can be obtained from
\begin{align}
	0 = \frac{\delta \Gamma_\text{2PI}}{\delta \phi (x)};\,\,\, 
	0 = \frac{\delta \Gamma_\text{2PI}}{\delta G_\varphi (x,y)};\,\,\,
	0 = \frac{\delta \Gamma_\text{2PI}}{\delta G_\chi (x,y)}.
\end{align}
Since we are interested in a spatially homogeneous system,
it is convenient to perform the spatial Fourier transformation.
Then, the Schwinger-Dyson eqs. for two-point correlators:
\begin{align}
	G^{\chi/\varphi^{-1}}_\text{0} \cdot G^{\chi/\varphi} = \delta_{\cal C} - \Pi^{\chi/\varphi} \cdot G^{\chi/\varphi}
	\label{eq:sde_on_c}
\end{align}
can be expressed as
\begin{align}
	\left[ \der_t^2 + \bm{p}^2 + M_{\chi/\phi}^2 (t) \right] G^{\chi/\varphi}_{J} (t,t';\bm{p})
	=& + i \int^t_{t'} d\tau\, \Pi^{\chi/\varphi}_J (t,\tau;\bm{p}) G_J^{\chi/\varphi} (\tau ,t';\bm{p})
	\label{eq:kbe_j} \\[5pt]
	\left[ \der_t^2 + \bm{p}^2 + M_{\chi/\phi}^2 (t) \right] G^{\chi/\varphi}_{H} (t,t';\bm{p})
	=& +i \int^t_{t_\text{ini}} d\tau\, \Pi^{\chi/\varphi}_J (t,\tau;\bm{p}) G_H^{\chi/\varphi} (\tau ,t';\bm{p}) \nn[5pt]
	& - i \int^{t'}_{t_\text{ini}} d\tau\, \Pi^{\chi/\varphi}_H (t,\tau;\bm{p}) G_J^{\chi/\varphi} (\tau,t';\bm{p})
	\label{eq:kbe_h}
\end{align}
where the effective masses are given by
\begin{align}
	M^2_\phi (t) & =  m_\phi^2 + \lambda^2 \int_{\bm{p}} G^\chi_H (t,t;\bm{p}), \\[5pt]
	M^2_\chi (t) &= m_{\chi,\text{th}}^2 + \lambda^2 \left[ \phi^2 (t) 
	+ \frac{1}{2} \int_{\bm{p}} G_H^\varphi (t,t';\bm{p}) \right].
\end{align}
Here we have explicitly written down the thermal mass of $\chi$ field, $m_{\chi,\text{th}}$,
that emerges from the gauge/Yukawa interaction with the background plasma,
and it is roughly evaluated as $m_{\chi,\text{th}} \sim gT (\sim yT)$ with $T$ being the temperature of background plasma.
Aside from the contribution of interactions with the background plasma,
the self energies of $\chi/\varphi$ are given by:
\begin{align}
	\Pi_{J}^{\chi}(x,y) \supset
	&+ 2 \lambda^4 \phi(x) \left[ G_J^\varphi (x,y) G_H^\chi (y,x) - G_H^\varphi (x,y) G_J^\chi (y,x) \right] \phi(y) \\
	&+ \lambda^4 G^\varphi_J (x,y) G^\varphi_H (x,y) G^\chi_H (y,x) 
	- \fr{\lambda^4}{2} \left[ G_H^{\varphi^2} (x,y) + G_J^{\varphi^{2}} (x,y) \right] G_J^\chi (y,x),  \\[10pt]
	\Pi_{H}^{\chi}(x,y) \supset&
	+ 2 \lambda^4 \phi(x) \left[ G_H^\varphi (x,y) G_H^\chi (y,x) - G_J^\varphi (x,y) G_J^\chi(y,x) \right] \phi(y) \\
	&- \lambda^4 G_J^\varphi (x,y) G_H^\varphi (x,y) G^\chi_J (y,x)
	+ \fr{\lambda^4}{2} \left[ G_H^{\varphi^2} (x,y) + G_J^{\varphi^{2}} (x,y) \right] G_H^\chi (y,x), \\[10pt]
	\Pi_{J}^{\varphi}(x,y) \supset& 
	+ 2\lambda^4 \phi (x) \left[ G_J^\chi (x,y) G_H^\chi (y,x) - G_H^\chi (x,y) G_J^\chi (y,x) \right] \phi(y) \\
	&+
	\lambda^4  \left[ G^\chi_J (x,y) G^\chi_H (y,x) - G^\chi_H (x,y) G^\chi_J (y,x) \right] G^\varphi_H (x,y) \nn
	&+ \lambda^4 \left[ G^\chi_H (x,y) G^\chi_H (y,x) - G^\chi_J (x,y) G^\chi_J (y,x) \right] G^\varphi_J (x,y), \\[10pt]
	\Pi_{H}^{\varphi}(x,y) \supset &
	+ 2 \lambda^4 \phi(x) \left[ G^\chi_H (x,y) G^\chi_H (y,x) - G^\chi_J (x,y) G^\chi_J (y,x) \right] \phi(y) \\
	&+ \lambda^4\left[ G^\chi_J (x,y) G^\chi_H (y,x) - G^\chi_H (x,y) G^\chi_J (y,x) \right] G^\varphi_J (x,y) \nn
	&+ \lambda^4 \left[ G^\chi_H (x,y) G^\chi_H (y,x) - G^\chi_J (x,y) G^\chi_J (y,x) \right] G^\varphi_H (x,y).
\end{align}

On the other hand,
the equation of motion for mean field is given by
\begin{align}
	0 &=  \left[ \der_t^2 + 3 H \der_t + M_\phi^2 (t)  \right] \phi(t) 
	- i \int^t_{t_\text{ini}} d\tau\, \Pi_J (t,\tau) \phi (\tau)
\end{align}
where 
\begin{align}
	\Pi_J (t,t') =
	\mathscr{F}_{\bm{x-y}} \star 
	&\lambda^4 \left[ G^\chi_H (x,y) G^\chi_H (y,x) G_J^\varphi (x,y)
	- G^\chi_J (x,y) G^\chi_J (y,x) G_J^\varphi (x,y)\right. \nn
	&\left.+ G^\chi_J (x,y) G^\chi_H (y,x) G_H^\varphi (x,y)
	- G^\chi_H (x,y) G^\chi_J (y,x) G_H^\varphi (x,y)  \right] (t,t';\bm{0}) \\[5pt]
	=
	\mathscr{F}_{\bm{x-y}} \star &\left[  \Pi^\varphi_J (x,y)|_{\phi=0} \right] (t,t;\bm{0})
\end{align}
with $\mathscr{F}_{\bm{x-y}}$ being the spacial Fourier transformation with respect to 
$\bm{x-y}$.
Note that here the adiabatic expansion of the Universe is taken into account explicitly.\footnote{
	Since we consider the regime where the cosmic expansion is adiabatic,
	it only red-shifts the particle distribution imprinted in Eqs.~\eqref{eq:kbe_j} and
	\eqref{eq:kbe_h}.
	Hence, we do not explicitly write down the effects of expanding background
	in Eqs.~\eqref{eq:kbe_j} and \eqref{eq:kbe_h}.
	We can reformulate the Schwinger-Dyson eqs.\ on CTP in terms of conformal time
	~\cite{Tranberg:2008ae}.
}

\subsection{Non-perturbative particle production}
\label{sec:np_prod}

First, let us study the non-perturbative $\chi$ particle production.
Obviously, in this case, the $\chi$ propagators are dynamical and hence
we have to study the evolution of $\phi$ and $G_\chi$ at least simultaneously.

To illustrate the essential feature, let us consider the following set of equations at first
discarding the self energy contributions~\cite{Berges:2004yj}:
\begin{align}
	0 &= \left[ \der_t^2 + m_\phi^2 \right] \phi (t), \\
	0 &= \left[ \der_t^2 +\bm{p}^2 + m_{\chi,\text{th}}^2 + \lambda^2 \phi^2 (t) \right] G^\chi_{J/H}(t,t';\bm{p}).
	\label{eq:kbe_apprx}
\end{align}
The applicability of these approximated equations is discussed later.
Since we neglect the finite density correction including the back-reaction to the oscillating scalar,
the first equation reads $\phi (t) = \tilde \phi \cos [m_\phi t]$.
Here we take the initial time as $t_\text{ini} = 0$ without loss of generality and
concentrate on the case $\lambda \tilde \phi \gg m_\phi$ in the following.
Then, let us turn to the latter equation.
Initially, the $\chi$ particles are assumed to be absent, so the initial condition for $G_H^\chi$
is given by
\begin{align}
	\left. G^\chi_H (t,t';\bm{p}) \right|_{t,t' = 0}
	&= \fr{1}{\sqrt{\bm{p}^2 + m_{\chi,\text{th}}^2}}, \\
	\der_t \der_{t'}\left. G^\chi_H (t,t';\bm{p}) \right|_{t,t' = 0}
	&= \sqrt{\bm{p}^2 + m_{\chi,\text{th}}^2}, \\
	\der_t \left. G^\chi_H (t,t';\bm{p}) \right|_{t,t' = 0}
	&= \der_{t'} \left. G^\chi_H (t,t';\bm{p}) \right|_{t,t' = 0} = 0.
\end{align}
Note that $G^\chi_J$ satisfies the canonical commutation relations:
\begin{align}
	\left. G^\chi_J (t,t';\bm{p}) \right|_{t' = t} &= \der_t \der_{t'} \left. G^\chi_J (t,t';\bm{p}) \right|_{t' = t} = 0, \\
	\der_{t'} \left. G^\chi_J (t,t';\bm{p}) \right|_{t' = t} &=  - \der_{t} \left. G^\chi_J (t,t';\bm{p}) \right|_{t' = t} 
	= i.
\end{align}
One finds the following factorized solutions:
\begin{align}
	G^\chi_H (t,t';\bm{p}) &= \left[ f_{\bm{p}}(t) f^\ast_{\bm{p}}(t') + f^\ast_{\bm{p}}(t) f_{\bm{p}}(t') \right], \\
	G^\chi_J (t,t';\bm{p}) &= \left[ f^\ast_{\bm{p}} (t) f_{\bm{p}} (t') - f_{\bm{p}} (t) f^\ast_{\bm{p}} (t') \right],
\end{align}
where the equation of motion for each mode is given by
\begin{align}
	0 = \left[ \der_t^2 + \bm{p}^2 + m_{\chi,\text{th}}^2 + \lambda^2 \phi^2 (t) \right] f_{\bm{p}} (t),
\end{align}
with the initial condition being
\begin{align}
	f_{\bm{p}} (0) = \fr{1}{\sqrt{2} \left[ \bm{p}^2 + m_{\chi,\text{th}}^2 \right]^{1/4}};\,\,\,\,\,
	\dot f_{\bm{p}} (0) = - i \fr{\left[ \bm{p}^2 + m_{\chi,\text{th}}^2 \right]^{1/4}}{\sqrt{2}}.
\end{align}
This is nothing but the \textit{Mathieu} equation.

Since the dispersion relation of $\chi$ particles depends on the oscillating scalar
$\omega_\chi^2 = \bm{p}^2 + m_{\chi,\text{th}}^2 + \lambda^2 \phi^2 (t)$,
the adiabaticity for $f_{\bm{p}}$  can be broken down $|\dot \omega_\chi/\omega_\chi^2| \gg 1$ 
when the $\phi$ passes through its potential origin $\phi \sim 0$.
The amplitude of mode function $f_{\bm{p}}$ suddenly grows at $\phi \sim 0$
and the $\chi$ particles are non-perturbatively produced consequently,
if the adiabaticity is broken down and 
the amplitude of $\phi$ is large enough $\lambda \tilde \phi \gg m_\phi$
as extensively studied in Refs.~\cite{Kofman:1994rk}.
This condition implies the following criteria for the non-perturbative 
production~\cite{Mukaida:2012qn,Mukaida:2012bz}:
\begin{align}
	\lambda \tilde \phi \gg \max \left[ m_\phi, \fr{m_{\chi,\text{th}}^2}{m_\phi} \right].
	\label{eq:np_cond_1}
\end{align}
If the condition Eq.~\eqref{eq:np_cond_1} is met, then 
the distribution function of $\chi$ particles suddenly acquires the order one value
after the first passage of non-adiabatic region: 
$|\phi| < \phi_\text{NP} := [m_\phi \tilde \phi / \lambda]^{1/2}$.
Here and hereafter we define the distribution function of particles $(\bullet = \varphi, \chi)$ as 
$f_{\bm{p}}^\bullet = [\der_t \der_{t'} + \Omega^{\bullet^2}_{\bm p}]G_H^\bullet (t,t';\bm{p})/(4 \Omega^\bullet_{\bm{p}})|_{t' \to t} - 1/2$
outside the non-adiabatic region: $\phi_\text{NP} \ll |\phi (t)|$~\cite{Boyanovsky:2004dj,Hamaguchi:2011jy}.\footnote{
	There are some ambiguities on the definition of particle number in terms of Green function~\cite{Garbrecht:2002pd}.
}
Here dispersion relations are defined as $\Omega_{\bm{p}}^\bullet := [M_\bullet^2 + \bm{p}^2]^{1/2}$
with $\bullet = \varphi,\chi$.
Then, the number density of $\chi$ particles is given by
\begin{align}
	n_\chi \simeq N_\text{d.o.f.} 
	\times \frac{k_\ast^3}{4\pi^3};~~~k_\ast :=  \left[ \lambda \tilde \phi m_\phi \right]^{1/2},
\end{align}
where $N_\text{d.o.f.}$ stands for the number of $\chi$ particles normalized by one complex scalar.

There are three remarks: 
\begin{itemize}
	\item First, 
	Eq.~\eqref{eq:np_cond_1} implies that the non-perturbative production is suppressed
	for $k_\ast \lesssim m_{\chi,\text{th}}$. 
	\item Second, here in Eq.~\eqref{eq:kbe_apprx}, we have neglected
	the dissipative effects of $\chi$ from background plasma imprinted in self energy of $\chi$
	[See for instance Eqs.~\eqref{eq:kbe_j} and \eqref{eq:kbe_h}].
	Let us estimate whether or not this effect disturbs the non-perturbative production.
	Though it is rather subtle to estimate the dissipative effects inside the non-adiabatic
	region since we cannot define $\chi$ particles,
	nevertheless we may roughly evaluate it as follows.
	If $\Gamma_\text{int} \delta t_\text{NP} \ll 1$ is satisfied with $\delta t_\text{NP} \sim k_\ast^{-1}$
	being the time scale which $\phi$ takes to pass the non-adiabatic region  and
	$\Gamma_\text{int}$ being the typical interaction rate of $\chi$ with background plasma,
	then we expect that the dissipation cannot disturb the non-perturbative production.
	In most cases we expect $\Gamma_\text{int} < gT$ and hence
	Eq.~\eqref{eq:np_cond_1} implies 
	$\Gamma_\text{int} \delta t_\text{NP} < m_{\chi,\text{th}} / k_\ast \ll 1$.
	\item Third, after the ``first'' passage of non-adiabatic region:
	$\phi_\text{NP} > |\phi (t)|$, the subsequent evolution crucially depends on the property of $\chi$.
	Here we assume that the $\chi$ can decay into other light particles with the rate being
	$\Gamma_{\chi,\text{dec}} \sim y^2 m_{\chi} (|\phi (t)|)$,
	which is imprinted in the $\chi$'s self energy in Eqs.~\eqref{eq:kbe_j} and \eqref{eq:kbe_h}.
	Note that since the decay is dominated at the outside of non-adiabatic region, 
	the concept of $\chi$ particle decay is justified a posteriori.

	If the decay rate is so large that $y^2 \lambda \tilde \phi \gg m_\phi$,
	the non-perturbatively produced $\chi$ can decay completely well before the $\phi$
	moves back to its origin~\cite{Felder:1998vq}. 
	In this case, the dissipation rate reads
	\begin{align}
			\Gamma_\phi^\text{NP} \sim N_\text{d.o.f} \times
			 \frac{\lambda^2 m_\phi}{2 \pi^4 |y|}. 
	\end{align}

	On the other hand, for $y^2 \lambda \tilde \phi \ll m_\phi$ (or stable $\chi$),
	the parametric resonance occurs due to the induced emission factor of previously produced
	$\chi$ particles.
	Then, the key assumption that the background plasma can remain in thermal equilibrium
	becomes questionable.
	Thus, we may have to follow the dynamics of these variables in background plasma
	simultaneously.
	We do not consider this case in the following for simplicity.
	(See {\it e.g.} Ref.~\cite{Moroi:2013tea}.)
\end{itemize}

\subsection{Coarse-Grained eqs. for the mean field} 
\label{sec:cgeq_mf}

It is practically difficult to follow the evolution of mean field $\phi$ and propagators $G_{\varphi/\chi}$ completely.
See Refs.~\cite{Berges:2002cz} for recent developments to study the dynamics of oscillating scalar from first principles.

However, there are particular cases where the equations become rather simple;
we do not have to track the evolution of $\chi$ propagators in the following cases in contrast to the case studied in Sec.~\ref{sec:np_prod}.
(a) If the oscillation of the scalar field is so slow that even $\chi$ particles can regard the scalar condensation as a
static background, 
one can reduce the full set of equations to the coarse-grained equations with
assuming that low order correlators of fast $\chi$ fields can be approximated with the thermal ones.
(b) If the amplitude of oscillating scalar is smaller than the thermal mass of $\chi$ $\lambda \tilde \phi \ll m_{\chi,\text{th}}$,
one can compute the thermal corrections by simply assuming that the background plasma including $\chi$ particles
remains in thermal equilibrium.

Let us discuss these two cases in the following.

\subsubsection{Slowly oscillating scalar}
\label{sec:slow}

Let us consider the case where the dynamics of scalar condensation is so slow that
the low order correlators of $\chi$ field can closely track the thermal equilibrium ones.
Hence, in the following, we concentrate on the region where the adiabaticity is not broken down
and the oscillation time scale is much slower than the typical interaction time scale of light particles.	
Interestingly, the condition that the non-perturbative production does not occur $(k_\ast^2 \ll m_{\chi,\text{th}}^2)$
[See Eq.~\eqref{eq:np_cond_1}] implies that there is enough time for
the $\chi$ particles to become as abundant as thermal ones when the $\phi$ passes through the region $|\phi (t)| < T/\lambda$.
In this case, the obtained set of equations can be reduced to coarse-grained equations.

In the following, we focus on the regime $\lambda |\phi (t)| < T$\footnote{
	A field value $\phi (t)$ should not be confused with the amplitude $\tilde \phi$.
	The amplitude $\tilde \phi$ is not necessarily small enough to satisfy $\lambda \tilde \phi < T$.
} where the $\chi$ particles are relativistic. 
This is because, as shown in Sec.~\ref{sec:dyn}, this regime dominates the oscillation-averaged dissipation rate
of $\phi$ in most cases.

First, we evaluate the effective mass term $M_\phi^2 (t)$.
Since the dynamics of $\phi$ field is slow, one can approximate the Eq.~\eqref{eq:sde_on_c} around
a time $\bar t$ with $\phi(\bar t) = \bar \phi$:
\begin{align}
	G_\chi (t,t';\bm{p}) =& \left. G_{\chi,\text{th}} (t,t';\bm{p}) \right|_{\bar \phi}
	- i \int_{\cal C} d\tau d\tau'\, \left. G_{\chi,\text{th}} (t,\tau;\bm{p}) \right|_{\bar \phi}
	K (\tau,\tau'; \bar \phi) G_\chi (\tau,t';\bm{p})\\
	=& \left. G_{\chi,\text{th}} (t,t';\bm{p})  \right|_{\bar \phi}
	- 2 i  \lambda^2 \bar\phi \int_{\cal C} d\tau\, \left. G_{\chi,\text{th}} (t,\tau;\bm{p}) \right|_{\bar \phi}
	\delta \phi (\tau;\bar t) \left.G_{\chi,\text{th}} (\tau,t';\bm{p}) \right|_{\bar \phi}
	\label{eq:g_app_chi} \\
	&+ \cdots, \nonumber
\end{align}
where $K(\tau, \tau'; \bar \phi) := \lambda^2 \left[ 2\bar \phi \delta \phi (\tau; \bar t) + \delta \phi^2 (\tau;\bar t) 
\right] \delta_{\cal C} (\tau, \tau') - i \left[ \Pi_\chi - \Pi_\chi |_{\bar \phi} \right] (\tau,\tau')$.
Here $G_{\chi,\text{th}}|_{\bar \phi}$ is the thermal propagators with $\phi (\bar t) = \bar \phi$
and we implicitly assume that the average of arguments in the Green function is near $\bar t$: 
$(t+t')/2 \simeq \bar t$.
In the second equality, we have neglected contributions in $K$ from higher orders in $\delta \phi$ and also 
from the self energy of $\chi$.\footnote{
	In the large field value regime $\lambda |\phi (t)| \gg T$, 
	the contribution from $\chi$'s self energy may dominate
	the dissipation factor encoded in the effective mass term $M_\phi^2 (t)$.
}
Note that in estimating the dissipation rate of slowly oscillating scalar,
the latter self energy contribution can be comparable to the result from $\lambda^2 \bar \phi \delta \phi$,
and the resultant dissipation rate can change by several factors~\cite{Jeon:1994if,BasteroGil:2010pb}.
Nevertheless, we roughly estimate the dissipation factor with dropping contributions from the self energy
without taking care of factor uncertainties.

By inserting Eq.~\eqref{eq:g_app_chi} to the effective mass term $M_\phi^2 (t)$,
one finds
\begin{align}
	M^2_\phi(\bar t) = m^2_\phi &+ \lambda^2 \int_{\bm{p}}
	\left.G_H^{\chi,\text{th}} (\bar t, \bar t;\bm{p}) \right|_{\bar \phi} \label{eq:th_mass_phi} \\ 
	& - 4 i  \lambda^4 \bar\phi
	\int_{\cal C} d\tau \int_{\bm{p}}  \left. G_{\chi,\text{th}} (\bar t,\tau;\bm{p}) \right|_{\bar \phi}
	\delta \phi (\tau;\bar t) \left.G_{\chi,\text{th}} (\tau,\bar t;\bm{p}) \right|_{\bar \phi} + \cdots. \label{eq:diss_phi}
\end{align}
Note that the contributions from thermal log and possible Coleman-Weinberg potentials may be imprinted in Eq.~\eqref{eq:th_mass_phi}.
For $\lambda |\phi| < T$, the former term [Eq.~\eqref{eq:th_mass_phi}] encodes the thermal mass of $\phi$
from the $\chi$ particles $m_{\phi,\text{th}}$ [Eq.~\eqref{eq:th_mass}]:
\begin{align}
	m_{\phi,\text{th}}^2 
	= N_\text{d.o.f.} \times \lambda^2 \int_{\bm p} 
	\frac{f_\text{B}(|{\bm p}|)}{|{\bm p}|} = N_\text{d.o.f.} \times \frac{\lambda^2 T^2}{12}
	\label{eq:th_mass_2}
\end{align}
at the leading order in high temperature expansion.
The latter term [Eq.~\eqref{eq:diss_phi}] encodes the friction term of $\phi$ due to the
abundant $\chi$ particles, so let us concentrate on this term:
\begin{align}
	- 4 i \lambda^4 \bar \phi \int_{t_\text{ini}}^{\bar t} d\tau\,  \delta\phi (\tau;\bar t) \int_{\bm{p}}
	\left[  G_>^{\chi,\text{th}} (\bar t,\tau;\bm{p}) |_{\bar \phi}  G_<^{\chi,\text{th}} (\tau,\bar t;\bm{p}) |_{\bar \phi} 
	-  G_<^{\chi,\text{th}} (\bar t,\tau;\bm{p}) |_{\bar \phi}  G_>^{\chi,\text{th}} (\tau, \bar t;\bm{p}) |_{\bar \phi} 
	\right]. \nonumber
\end{align}
Hereafter, we adopt the following approximations: $\delta \phi (\tau; \bar t) = \dot \phi (\tau - \bar t) + \cdots$ and
$t_\text{ini} \to - \infty$ since the motion of $\phi$ is assumed to be sufficiently slow. Then, the friction term
of $\phi$ reads
\begin{align}
	\Gamma_\phi^{(a)} &= - 4 i \lambda^4 \phi (\bar t)^2
	\int^0_{- \infty} d\tau\, \tau
	\int_{\omega_1, \omega_2,\bm{p}}
	e^{i (\omega_1 - \omega_2) \tau}
	\left[ f_\text{B} (\omega_1) - f_\text{B} (\omega_2) \right]
	\rho_{\chi,\text{th}} (\omega_1,\bm{p}) |_{\bar \phi}\, \rho_{\chi,\text{th}} (\omega_2,\bm{p}) |_{\bar \phi} \nn
	&= 
	\frac{2\lambda^4 \phi^2 (\bar t)}{T} \int_{\omega, \bm{p}} 
	\left[ - f_\text{B} (\omega) f_\text{B} (-\omega) \right] 
	\rho_{\chi,\text{th}} (\omega, \bm{p})\rho_{\chi,\text{th}} (\omega, \bm{p}). \label{eq:diss_zero_m}
\end{align}
Here we have used the relations~Eqs.~\eqref{eq:rel_kms_1} and \eqref{eq:rel_kms_2}: 
$G_>^\text{th} (P) = [1 + f_\text{B} (p^0)] \rho_{\chi,\text{th}} (P)$,
$G^\text{th}_< (P) = f_\text{B} (p^0) \rho_{\chi,\text{th}} (P)$ with the spectral density being $
\rho_{\chi,\text{th}} (P) = G_J^{\chi,\text{th}} (P)$.
Assuming the Breit-Wigner form for the spectral density of $\chi$ quasi-particles
\begin{align}
	\rho_{\chi,\text{th}} (P) \simeq
	\frac{2 p_0 \Gamma_{\bm{p}}}{[p_0^2 - \Omega_{\bm{p}}^2]^2 +  [p_0 \Gamma_{\bm{p}}]^2}
\end{align}
and roughly approximating the thermal width as $\Gamma \sim \alpha T^2/\omega$,
one can estimate the dissipation factor of $\phi$ as
\begin{align}
	\Gamma_\phi^{(a)} \sim N_\text{d.o.f.} \times \frac{\lambda^4 \phi^2 (\bar t)}{\pi^2 \alpha T}.
	\label{eq:diss_ann}
\end{align}

Second, let us evaluate the non-local term: $-i\Pi_J \ast \phi$.
Again we assume that the dynamics of $\phi$ is sufficiently slow
$\delta \phi (\bar t ;\tau)=\dot \phi (\tau - \bar t) + \cdots$ and $t_\text{ini} \to - \infty$,
and hence the non-local term can be approximated with
\begin{align}
	-i \int^{\bar t}_{-\infty} d\tau\, \Pi_J (\bar t,\tau) \phi (\tau) 
	\supset - i \int^{\bar t}_{-\infty}  d\tau\, \left. \Pi_J (\bar t-\tau;\bar t) \right|_{\bar \phi} \delta \phi (\tau;\bar t)
	\simeq \dot \phi (\bar t) 
	\left[\frac{\left.\Pi_J (\omega;\bar t) \right|_{\bar \phi} }{ 2 \omega} \right]_{\omega \to 0}
\end{align}
where the self energy is approximated by the
gradient-expansion: $\Pi_J (\bar t, \tau) = \Pi_J (\bar t - \tau;[\bar t + \tau]/2)
=\Pi_J (\bar t - \tau; \bar t) + \cdots$.\footnote{
Note that the correlation of two distinct time in the self energy is expected to decay much faster than
the motion of oscillating scalar since the scalar oscillates much slower than the typical interaction in 
thermal plasma.
}
Here we have extracted a term that contributes to the dissipation rate.
Thus, the dissipation rate reads
\begin{align}
	\Gamma_\phi^{(b)} 
	= \left[ \frac{\left. \Pi_J (\omega; \bar t) \right|_{\bar \phi}}{2 \omega} \right]_{\omega \to 0}.
	\label{eq:diss_scat}
\end{align}
Since there are no $\varphi$ particles initially, let us assume that the propagators of $\varphi$ can be approximated
with the vacuum one.
Then, one finds
\begin{align}
	\Pi_J (\omega;\bar t) |_{\bar \phi} =
	\lambda^4 \int_{K,Q,L} &\left[
	\( 1 + 2 f_{\text{B},k_0} \) \( 1 + 2 f_{\text{B},q_0} \) + 1 + 
	2\( 1 + 2 f_{\text{B},k_0} \) \text{sgn}(l_0)
	\right] \nn
	&\rho_{\chi,\text{th}} (K) |_{\bar\phi}\, \rho_{\chi,\text{th}} (Q) |_{\bar\phi}\,
	\rho_{\varphi,\text{vac}} (L)
\end{align}
where
\begin{align}
	\int_{K,Q,L} = \int \frac{d^4 K}{(2\pi)^4}\frac{d^4 Q}{(2\pi)^4}\frac{d^4 L}{(2\pi)^4}\,
	(2\pi)^4 \delta \( - \omega + k_0 + q_0 + l_0 \) 
	\delta \( \bm{k} + \bm{q} + \bm{l} \).
\end{align}
Although, in general, the distribution function of $\phi$ depends on the time $\bar t$,
as a reference point, we estimate the dissipation factor of $\phi$ in the case where
there is no $\phi$ particles.
Then the self energy can be evaluated as
\begin{align}
	\Pi_J (\bar t; \omega)|_{\bar \phi} \simeq
	N_\text{d.o.f.} \frac{\lambda^4 \omega}{4\pi^3}
	\(
	\int_{m_{\chi,\text{th}}} d\Omega \left[ - f_\text{B} (\Omega) f_\text{B} (- \Omega) \right]
	2 \frac{\Omega^2 - m_{\chi,\text{th}}^2}{m_{\chi,\text{th}}^2} \).
\end{align}
Here we have dropped the Boltzmann suppressed term for brevity.
Thus, the dissipation rate reads
\begin{align}
	\Gamma_\phi^{(b)} \sim N_\text{d.o.f.} \times \frac{\lambda^4 T^3}{12 \pi  m_{\chi,\text{th}}^2}.
	\label{eq:diss_scat_2}
\end{align}

\subsubsection{Oscillating scalar with small amplitude} 
\label{sec:cgeq_mf_small}

Then, let us consider the case where the amplitude $\tilde \phi$ is smaller than the thermal mass of $\chi$
$\lambda \tilde \phi \ll gT$, but not necessarily $m_\phi \ll \alpha T$ 
in contrast to the previous Sec.~\ref{sec:cgeq_mf}.
In this case, we expect the background plasma including $\chi$ particles
to remain in thermal equilibrium during the course of $\phi$'s oscillation for the following reasons.
First, the $\phi$-dependent mass term of $\chi$ can be safely neglected.
Second, the energy transportation time scale from the oscillating scalar $\phi$ to thermal plasma 
is much slower than the typical interaction time scale of thermal plasma in our case.
Third,
the broad resonance does not occur in this case since $\lambda \tilde \phi \ll m_{\chi,\text{th}}$ violates the condition 
for non-perturbative particle production: Eq.~\eqref{eq:np_cond_1}.
The narrow resonance also does not occur in our cases.
This is because
at least $m_\phi > m_{\chi,\text{th}}$ is required for the narrow resonance to take place, and in addition
the growth rate of narrow resonance should be larger than the decay and dissipation rate of $\chi$~\cite{Kasuya:1996np}:
$\lambda^2 \tilde \phi^2 / m_\phi \gg \max[\alpha T, y^2 \lambda \tilde \phi]$
in order for the induced emission to be efficient.
These conditions are unlikely to be satisfied in most cases of our interest.
(See the discussion in Appendix.~\ref{sec:narrow}.)
Therefore, one can calculate thermal corrections to oscillating scalar field by simply
assuming that the background plasma including $\chi$ particles can remain in thermal equilibrium
\cite{Yokoyama:2004pf,Drewes:2010pf}.

First, we evaluate the effective mass term $M_\phi^2 (t)$.
We follow the arguments in Ref.~\cite{Yokoyama:2004pf}.
Since the amplitude is small compared to the thermal mass of $\chi$,
the approximate solution of $\chi$'s propagator can be obtained in the similar way as the former section:
\begin{align}
	G_\chi (t,t';\bm{p}) &=
	- i \int_{\cal C} d\tau d\tau'\, \left. G_{\chi,\text{th}} (t,\tau;\bm{p}) \right|_{\bar \phi}
	K' (\tau,\tau') G_\chi (\tau,t';\bm{p})\\
	&= -  i  \lambda^2  \int_{\cal C} d\tau\, \left. G_{\chi,\text{th}} (t,\tau;\bm{p}) \right|_{\bar \phi=0}
	\phi^2(\tau ) \left.G_{\chi,\text{th}} (\tau,t';\bm{p}) \right|_{\bar \phi=0} + \cdots,
	\label{eq:g_app_chi_2}
\end{align}
where $K'(\tau, \tau') := \lambda^2 \phi^2 (\tau) \delta_{\cal C} (\tau, \tau') - i \left[ \Pi_\chi - \Pi_\chi |_{\bar \phi = 0} \right] (\tau,\tau')$.
In the following, we drop the subscript $\bar \phi = 0$ for brevity.
Aside from the thermal mass of $\phi$,
this term leads to the following contribution:
\begin{align}
	&-2 i \lambda^4 \int^{\bar t}_{-\infty} d\tau\, \phi^2 (\tau) 
	\int_{\bm{p}}
	\left[  G_>^{\chi,\text{th}} (\bar t,\tau;\bm{p})  G_<^{\chi,\text{th}} (\tau,\bar t;\bm{p})
	-  G_<^{\chi,\text{th}} (\bar t,\tau;\bm{p})  G_>^{\chi,\text{th}} (\tau, \bar t;\bm{p})
	\right] \nn
	=& - 2i \lambda^4 \int^{\infty}_{0} d\tau\, \phi^2 (\bar t -\tau) \frac{\der}{\der \tau} D(\tau) \nn
	=& + 2i \lambda^4 \phi^2 (\bar t) D(0) - 4 i \lambda^4 \int^\infty_0 d \tau\, 
	\phi (\bar t -\tau) \dot \phi (\bar t -\tau) \int_\omega e^{-i\omega \tau} D(\omega) \label{eq:diss_phi_dep_s}
\end{align}
where
\begin{align}
	\frac{\der}{\der \tau} D(\tau) := 
	\int_{\bm{p}}
	\left[  G_>^{\chi,\text{th}} (\tau;\bm{p})  G_<^{\chi,\text{th}} (-\tau;\bm{p})
	-  G_<^{\chi,\text{th}} (\tau;\bm{p})  G_>^{\chi,\text{th}} (-\tau;\bm{p})
	\right].
\end{align}
This implies
\begin{align}
	- i \omega D(\omega) = \int d\tau\, e^{i \omega \tau}
	\int_{\bm{p}}
	\left[  G_>^{\chi,\text{th}} (\tau;\bm{p})  G_<^{\chi,\text{th}} (-\tau;\bm{p})
	-  G_<^{\chi,\text{th}} (\tau;\bm{p})  G_>^{\chi,\text{th}} (-\tau;\bm{p})
	\right].
\end{align}
Let us concentrate on the latter term of Eq.~\eqref{eq:diss_phi_dep_s} that encodes the dissipation rate.
Inserting $\phi (\bar t - \tau) = \phi (\bar t) \cos [M_{\phi}^\text{L} \tau] - [\dot \phi (\bar t)/M_{\phi}^\text{L}] 
\sin [M_{\phi}^\text{L} \tau ]$, one finds
\begin{align}
	&- 4 i \lambda^4 \int^\infty_0 d\tau \int_\omega  e^{- i \omega \tau} D (\omega)
	\left[ \phi (\bar t) \dot \phi (\bar t) \cos \( 2 M_\phi^\text{L} \tau \)
	- \frac{1}{2 M_\phi^\text{L}} \( \dot \phi^2 (\bar t) - M_\phi^{\text{L}^2} \phi^2 (\bar t) \)
	\right] \nn
	=: & + D_1 \phi (\bar t) \dot \phi (\bar t) + D_2 
	\frac{1}{2 M_\phi^\text{L}} \( \dot \phi^2 (\bar t) - M_\phi^{\text{L}^2} \phi^2 (\bar t) \).
\end{align}
Here the leading contribution to the effective mass of $\phi$ is denoted by 
$M_\phi^\text{L} = [m_\phi^2 + m_{\phi,\text{th}}^2]^{1/2}$.
Since the second term vanishes if we consider the oscillation time averaged evolution equation of 
$\phi$'s energy density,
we concentrate on the first term $D_1$ that leads to the dissipation of oscillating scalar.
By definition, we have
\begin{align}
	D_1 =& - 4 i \lambda^4 \int^\infty_0 d\tau\, \int_\omega e^{- i \omega \tau} D(\omega)
	\cos \( 2M_\phi^\text{L} \tau \) \nn[5pt]
	=&
	\lambda^4 \frac{1}{M_\phi^\text{L}}
	\int d\tau\, e^{i 2 M_\phi^\text{L} \tau}
	\int_{\bm{p}}
	\left[  G_>^{\chi,\text{th}} (\tau;\bm{p})  G_<^{\chi,\text{th}} (-\tau;\bm{p})
	-  G_<^{\chi,\text{th}} (\tau;\bm{p})  G_>^{\chi,\text{th}} (-\tau;\bm{p})
	\right] \nn[5pt]
	=&
	2\lambda^4 \left[ \frac{1}{\omega} 
	\int_P [1 + f_\text{B} (p_0) + f_\text{B} (\omega - p_0)] \rho_{\chi,\text{th}} (p_0,\bm{p}) 
	\rho_{\chi,\text{th}} (\omega - p_0,\bm{p})
	\right]_{\omega=2M_\phi^\text{L}}.
\end{align}
Therefore, the dissipation rate can be expressed as
\begin{align}
	\Gamma\a_\phi = 
	2 \lambda^4 \phi^2 (\bar t)  \left[ \frac{1}{\omega} 
	\int_P [1 + f_\text{B} (p_0) + f_\text{B} (\omega - p_0)] \rho_{\chi,\text{th}} (p_0,\bm{p}) 
	\rho_{\chi,\text{th}} (\omega - p_0,\bm{p})
	\right]_{\omega=2M_\phi^\text{L}}.
\end{align}
Taking the vanishing effective mass limit $M_\phi^\text{L} \to 0$,
one can obtain Eq.~\eqref{eq:diss_zero_m} consistently.
In contrast to the former case [Eq.~\eqref{eq:diss_ann}], the effective mass of oscillating scalar
can be larger than the thermal mass of $\chi$: $m_\phi > m_{\chi,\text{th}}$.
Then, the perturbative decay (annihilation) of $\phi$ into two $\chi$ quasi-particles is kinematically allowed
and the dissipation factor is given by
\begin{align}
	\Gamma^{(a)}_\phi
	= N_\text{d.o.f.} \times
	\frac{\lambda^4 \phi^2 (\bar t)}{8 \pi m_\phi} \sqrt{1 - \frac{m_{\chi,\text{th}}^2}{m_\phi^2}} 
	\left[ 1 + 2 f_\text{B} (m_\phi) \right] \theta \( m_\phi^2 - m_{\chi,\text{th}}^2 \)
	\label{eq:diss_ann_s}
\end{align}
for $\lambda \tilde \phi < m_{\chi,\text{th}} < m_\phi$.

Next, let us evaluate the non-local term: $- i\Pi_J\ast \phi$.
A similar computation yields
\begin{align}
	-i \int^{\infty}_{0}  d\tau\, \Pi_J (\tau; \bar t)\phi (\bar t - \tau)
	\supset
	i\frac{\dot \phi (\bar t)}{M_\phi^\text{L}} \int_0^\infty d\tau\, \Pi_J (\tau;\bar t) \sin \( M_\phi^\text{L} \tau \)
	= \dot \phi (\bar t) \left[ \frac{\Pi_J (\omega;\bar t)}{2 \omega} \right]_{\omega = M_\phi^\text{L}}.
\end{align}
Here we have extracted a term that contributes to the dissipation factor.
Therefore, the dissipation rate is given by
\begin{align}
	\Gamma\b_\phi = \left[ \frac{\Pi_J (\omega;\bar t)}{2 \omega} \right]_{\omega = M_\phi^\text{L}}.
\end{align}
Obviously one can obtain the former result [Eq.~\eqref{eq:diss_scat}] by taking $M_\phi^\text{L} \to 0$.
In the former section case $M_\phi^\text{L} \ll \alpha T$, 
the phase space for the scattering $\phi \chi \to \varphi \chi$ is quite suppressed
since the final particles are almost collinear in the rest frame of thermal plasma.
For comparison, let us evaluate the dissipation rate in the case of $\alpha T \ll m_\phi \ll T$.
In this case, one finds
\begin{align}
	\Gamma\b_\phi &\sim N_\text{d.o.f.} \frac{\lambda^4}{2m_\phi} \frac{2}{(2\pi)^3} 
	\int d\Omega_k \, f_\text{B} (\Omega_k) \int^{\Omega_k} d\Omega_q \nn[5pt]
	&\simeq N_\text{d.o.f.}\times \frac{\lambda^4 T^2}{48 \pi m_\phi}
	\label{eq:diss_scat_s}
\end{align}
where there is no $\varphi$ particles.
If the $\varphi$ particles are as abundant as thermal ones, then the dissipation rate increases by a factor of $3/2$.

\subsection{Coarse-Grained eqs. for the $\varphi$ particles} 
\label{sec:cgeq_pt}

In this section, we discuss the evolution of $\varphi$'s propagators $G^{\varphi}_{J/H}$
assuming that the background plasma remains in thermal equilibrium.
Since these eqs.\ are coupled non-linear integro-differential equations,
they are practically more difficult to study than the Boltzmann equations.
As extensively studied previously (See {\it e.g.}\ 
Refs.~\cite{Kadanoff:1962,Calzetta:1986cq,Ivanov:1999tj,Prokopec:2003pj,Boyanovsky:2004dj,Berges:2004yj,Berges:2005md,
Hohenegger:2008zk,Anisimov:2008dz,Garbrecht:2011xw,Hamaguchi:2011jy,Drewes:2012qw}), 
to make the problem more tractable,
the Boltzmann-like equation are frequently derived from
the Kadanoff-Baym eqs.\ under several assumptions: Typically
(i) Quasi-particle spectrum, (ii) Separation of time scales, (iii) Negligence of finite time effects $t_\text{ini} \to - \infty$.

We consider the small amplitude regime, $\lambda \tilde \phi \ll gT$, for the following reasons.
Technically, within this regime, the background plasma including $\chi$ particles can be regarded as thermal bath
and the above assumptions are likely to be satisfied.\footnote{
	For instance, in the large amplitude regime, 
	we are interested in processes around the origin $\lambda |\phi (t)| \ll T$ where the $\chi$ becomes abundant.
	The time scale that the oscillating $\phi$ takes to pass this region is given by $\delta t \sim T/(\lambda m_\phi \tilde\phi)$.
	However, this time scale is too short to tell what is $\varphi$ particle because $m_\phi \delta t \ll 1$.
}
Practically, as shown in Sec.~\ref{sec:dyn}, 
the thermalization of $\varphi$ particles become important when the amplitude of oscillating scalar becomes 
smaller than the temperature of background plasma: $\tilde \phi < T$.

From the Kadanoff-Baym eqs.\ of $\varphi$ [Eqs.~\eqref{eq:kbe_j} and \eqref{eq:kbe_h}],
one can derive the Boltzmann-like equation for $\varphi$ particles under the above assumptions
and the adiabatic expansion of the Universe:
\begin{align}
	\der_t f_{\bm{p}}^\varphi - H \bm{p} \cdot \der_{\bm{p}} f_{\bm{p}}^\varphi = {\cal C} [f^\varphi_{\bm{p}}]
\end{align}
where
\begin{align}
	{\cal C} [f^\varphi_{\bm{p}}] 
	= \fr{i}{2 \Omega_{\bm{p}}} \int^t_{t_\text{ini}} d \tau
	\left[ \Pi_J^\varphi (t,\tau;\bm{p}) \der_t G^\varphi_H (\tau,t;\bm{p}) 
	- \Pi^\varphi_H (t,\tau;\bm{p}) \der_t G^\varphi_J (\tau ,t;\bm{p}) \right].
\end{align}
The collision terms with the on-shell approximation for $\chi$ quasi-particles are given by
\begin{align}
	\left. {\cal C} [f^\varphi_{\bm{p}}] \right|_\text{4pt} 
	= 8 N_\text{d.o.f.} \lambda^4 \int_{\bm{l,q,k}} & 
	(2\pi)^3 \delta (\bm{p-l-q-k}) 
	\fr{1}{2\Omega_{\bm{p}}^\varphi 2\Omega_{\bm{l}}^\varphi 2\Omega_{\bm{q}}^\chi 2\Omega_{\bm{k}}^\chi}
	\int^t_{t_\text{ini}} d\tau  
	\sum_{\{ s_i =\pm \}} \nn
	\Big\{ & s_1
	(\theta(s_1) + f^\varphi_{\bm{p}}) (\theta(s_2) + f^\varphi_{\bm{l}})  (\theta(s_3) + f^\chi_{\bm{q}})  (\theta(s_4) + f^\chi_{\bm{k}}) \nn
	&
	\cos \left[ \( s_1\Omega_{\bm{p}}^\varphi + s_2\Omega_{\bm{l}}^\varphi 
	+ s_3\Omega_{\bm{q}}^\chi + s_4 \Omega_{\bm{k}}^\chi \) \( t - \tau \) \right]
	\Big\},
\end{align}
\begin{align}
	\left. {\cal C} [f^\varphi_{\bm{p}}] \right|_\text{3pt} 
	= 4 N_\text{d.o.f.} \lambda^4 \phi(t) \int_{\bm{q,k}} & (2\pi)^3 \delta (\bm{p - q - k}) 
	\fr{1}{2\Omega_{\bm{p}}^\varphi 2\Omega_{\bm{q}}^\chi 2\Omega_{\bm{k}}^\chi}
	\int^t_{t_\text{ini}} d\tau  \sum_{\{ s_i = \pm \}} \nn
	\Big\{ & s_1
		(\theta(s_1) + f^\varphi_{\bm{p}}) (\theta(s_2) + f^\chi_{\bm{q}})  (\theta(s_3) + f^\chi_{\bm{k}}) \nn
		&\Bigl(  \phi(t) \cos \left[ \(s_4 M_\phi +  
		s_1 \Omega_{\bm{p}}^\varphi +  s_2\Omega_{\bm{q}}^\chi + s_3\Omega_{\bm{k}}^\chi \) \( t - \tau \) \right]   \nn
	& - s_4 \fr{\dot \phi(t)}{M_\phi}  \sin \left[ \( s_4M_\phi + s_1 \Omega_{\bm{p}}^\varphi +  s_2\Omega_{\bm{q}}^\chi 
	+ s_3\Omega_{\bm{k}}^\chi \) \( t - \tau \) \right]
	\Bigr)
	\Big\}.
\end{align}
Sending the initial time to remote past $t_\text{ini} \to - \infty$, one finds
\begin{align}
	\left. {\cal C}[f^\varphi_{\bm{p}}] \right|_\text{4pt} 
	=&  4 N_\text{d.o.f.} \lambda^4 \int_{\bm{l,q,k}} (2\pi)^4 \delta (\bm{p-l-q-k}) 
	\fr{1}{2\Omega_{\bm{p}}^\varphi 2\Omega_{\bm{l}}^\varphi 2\Omega_{\bm{q}}^\chi 2\Omega_{\bm{k}}^\chi} \nn
	& \Big\{2\left[ (1 + f^\varphi_{\bm{p}}) f^\varphi_{\bm{l}} ( 1 + f^\chi_{\bm{q}} ) f^\chi_{\bm{k}}
	- f^\varphi_{\bm{p}} ( 1+ f^\varphi_{\bm{l}} ) f^\chi_{\bm{q}} ( 1 + f^\chi_{\bm{k}}) \right] 
	\delta \( \Omega_{\bm{p}}^\varphi - \Omega_{\bm{l}}^\varphi + \Omega_{\bm{q}}^\chi - \Omega_{\bm{k}}^\chi \) \nn
	&+\left[ (1 + f^\varphi_{\bm{p}}) ( 1 + f^\varphi_{\bm{l}} )  f^\chi_{\bm{q}} f^\chi_{\bm{k}}
	- f^\varphi_{\bm{p}} f^\varphi_{\bm{l}} ( 1 + f^\chi_{\bm{q}} ) ( 1 + f^\chi_{\bm{k}}) \right]
	\delta\( \Omega_{\bm{p}}^\varphi + \Omega_{\bm{l}}^\varphi - \Omega_{\bm{q}}^\chi - \Omega_{\bm{k}}^\chi \) \Big\}, \label{eq:boltz_4pt} \\
	\left. {\cal C} [f^\varphi_{\bm{p}}] \right|_\text{3pt} 
	= & 2 N_\text{d.o.f.} \lambda^4 \phi^2(t) \int_{\bm{q,k}}  (2\pi)^4 \delta (\bm{p - q - k}) 
	\fr{1}{2\Omega_{\bm{p}}^\varphi 2\Omega_{\bm{q}}^\chi 2\Omega_{\bm{k}}^\chi} \nn
	&\Big\{ 2 \left[ (1 + f^\varphi_{\bm{p}}) (1 + f^\chi_{\bm{q}}) f^\chi_{\bm{k}}
	- f^\varphi_{\bm{p}} f^\chi_{\bm{q}} ( 1 + f^\chi_{\bm{k}} ) \right]
	\delta \( - M_\phi +  \Omega_{\bm{p}}^\varphi +  \Omega_{\bm{q}}^\chi - \Omega_{\bm{k}}^\chi \) \nn
	&- \left[ f^\varphi_{\bm{p}} (1 + f^\chi_{\bm{q}}) ( 1 +f^\chi_{\bm{k}} )
	- ( 1 + f^\varphi_{\bm{p}} ) f^\chi_{\bm{q}} f^\chi_{\bm{k}} \right] 
	\delta \( M_\phi + \Omega_{\bm{p}}^\varphi -  \Omega_{\bm{q}}^\chi - \Omega_{\bm{k}}^\chi \)
	\Big\}. \label{eq:boltz_3pt}
\end{align}
Here we have dropped the contribution proportional to $\phi \dot \phi$ which vanishes with the oscillation time average.

\section{Conclusions and Discussion} 
\label{sec:conc}

We have studied the dynamics of a scalar field with the Lagrangian (\ref{eq:setup}).
Although the scalar field has a $Z_2$ symmetry which ensures the stability of the scalar in the vacuum,
its energy can be dissipated through the scattering with $\chi$ particles in thermal background.
In order to deal with such effects, we have calculated the dissipation coefficient of the scalar field based on the CTP formalism.
It is found that the dissipative effect is so efficient that the energy density of the coherent oscillation can be reduced
to a cosmologically harmless level if the coupling $\lambda$ is larger than the critical value (\ref{lambda_c}).
It is understood intuitively: if the typical thermalization rate of $\phi$ is larger than the Hubble expansion rate,
$\phi$ is expected to be thermalized.
If this is the case, the final relic abundance of the scalar field is determined by the standard calculation of the thermal relic abundance.

Let us mention possible applications.
Such a scalar field with a large field value during inflation could be a candidate of the curvaton.
The large scale fluctuation imprinted in the curvaton energy density can be turned into that of the radiation,
if the curvaton coherent oscillation decays/dissipates.
Even if the perturbative decay is prohibited, as described above, thermal dissipation effects can dissipate the coherently oscillating scalar into the radiation.
In the $Z_2$-symmetric case, however, it is unlikely that the scalar field dominates the Universe before it is dissipated.
Typically, the fraction of the energy density of the coherent oscillation to the total energy density of the Universe is much smaller than unity unless the initial field value is very close to the Planck scale as long as we require that the scalar field is completely dissipated.
It would lead to too large non-Gaussianity if it is the dominant source of the curvature perturbation.
Therefore, it is difficult to explain the observed curvature perturbation of the Universe by the $Z_2$-symmetric scalar field
without producing too large non-Gaussianity.

\section*{Acknowledgment}

This work is supported by Grant-in-Aid for Scientific
research from the Ministry of Education, Science, Sports, and Culture
(MEXT), Japan, No.\ 21111006 (K.N.), and No.\ 22244030 (K.N.).
The work of K.M. and M.T. is supported in part by JSPS Research Fellowships
for Young Scientists.

\appendix

\section{Narrow resonance}
\label{sec:narrow}

\begin{figure}[t]
\begin{center}
\includegraphics[scale=0.6]{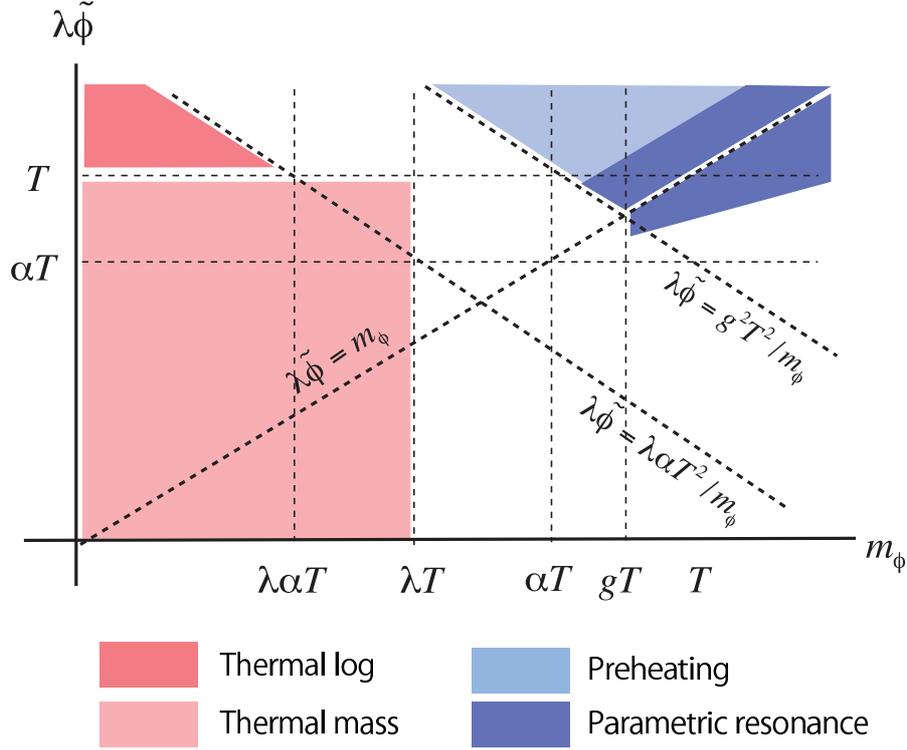}
\caption{ 
	Rough sketch for the classification of the scalar oscillation on $(m_\phi, \lambda\tilde\phi)$-plane.
	Light red: oscillation with thermal mass. Dark red: oscillation with thermal logarithm.
	Light blue: non-perturbative particle production occurs. Dark blue: parametric resonance occurs.
}
\label{fig:gamma}
\end{center}
\end{figure}

In this appendix, we comment on the effect of narrow resonance which could happen in some parameter ranges.
We follow the arguments in Refs.~\cite{Kofman:1994rk,Kasuya:1996np,Shtanov:1994ce}.

First of all, we recall that the resonance regime does not appear if the condition (\ref{eq:decay_cond}) is satisfied.
Combined with the condition for the non-perturbative particle production (\ref{eq:np_cond}),
the broad resonance occurs when
\begin{align}
	\fr{m_\phi}{y^2} \gg \lambda \tilde \phi \gg \max \left[ m_\phi, \fr{m_{\chi,\text{th}}^2}{m_\phi} \right].
\end{align}

On the other hand, the narrow resonance takes place for $q \equiv \lambda^2\tilde\phi^2/m_\phi^2 \ll 1$
and $m_\phi \gg gT$. Thus we concentrate on this case.
Hereafter we assume $y \sim g$ for simplicity.

Let us consider the first instability band for the $\chi$ at $k \simeq m_\phi$, where $k$ is the physical wavenumber of $\chi$
in the Fourier mode.
The width of the instability band is given by $\Delta k /k \sim q$, and the growth rate of $\chi$ is given by $\sim qm_\phi$.\footnote{
	This is understood as the perturbative decay of $\phi$ combined with the induced emission effect.
	The perturbative decay rate of $\phi$ is given by $\Gamma_\phi \sim \lambda^4\tilde\phi^2/m_\phi$
	and the phase space density of $\chi$ is given by $f_\chi(k) \sim n_\chi / (k^2\Delta k)$ peaked around $k \simeq m_\phi$.
	Thus the evolution of the number density is governed by $\dot n_\chi \sim \Gamma_\phi n_\phi f_\chi \sim q m_\phi n_\chi$.
	This gives $n_\chi \propto \exp (q m_\phi t)$.
}
In order for the resonance to occur, the momentum distribution of $\chi$ must not be disturbed in a time interval of $(qm_\phi)^{-1}$.
The sources for termination of the resonance are the $\chi$ interaction with thermal plasma 
and the decay of $\chi$.
Thus we need $q m_\phi \gg \max[\Gamma_\chi (\sim \alpha T), y^2 \lambda \tilde \phi]$ for the resonance.
Another source is the Hubble expansion, which redshifts the physical momentum of $\chi$.
The time required for removing $\chi$ particles from the resonance band $k \sim m(1 \pm q)$ is $\Delta t_H \sim q/H$.
During this time interval, the growth of $\chi$ number density is at most $\sim \exp(qm_\phi \Delta t_H) \sim \exp(q^2 m_\phi/H)$.
Therefore, we also need $q^2 m_\phi \gg H$ for the efficient resonance.
If these two conditions are satisfied, the $\chi$ number density exponentially grows due to the narrow resonance effect.
Fig.~\ref{fig:gamma} depicts the parameter region where the narrow resonance can occur.

If it happens, the end of the exponential growth may be caused by the self-interaction of $\chi$.
For example, the rate of the self-annihilation process $\chi\chi \to gg$ (gauge bosons) is estimated as
$\Gamma_{\chi\chi \to gg} \sim \alpha^2 n_\chi / m_\phi^2$.
If this becomes equal to $qm_\phi$, the resonance stops.
It happens at $n_\chi \sim (\lambda/\alpha)^2 n_\phi$. 
Therefore, for $\lambda < \alpha$, we have $\rho_\chi < \rho_\phi$ at the end of resonance
and hence it does not drastically affect the dynamics of $\phi$ field. 
(It is same order of the energy loss rate at the preheating stage just before the narrow resonance regime.)
The evolution of $\phi$ field after the end of the resonance should be solved in a way described in the text and the results are not much affected.




\begin{thebibliography}{99}



\bibitem{Coughlan:1983ci}
  G.~D.~Coughlan, W.~Fischler, E.~W.~Kolb, S.~Raby, G.~G.~Ross,
  Phys.\ Lett.\ B {\bf 131}, 59 (1983);
  J.~R.~Ellis, D.~V.~Nanopoulos, M.~Quiros,
  Phys.\ Lett.\ B {\bf 174}, 176 (1986);
  A.~S.~Goncharov, A.~D.~Linde, M.~I.~Vysotsky,
  Phys.\ Lett.\ B {\bf 147}, 279 (1984).
%



\bibitem{deCarlos:1993jw}
  B.~de Carlos, J.~A.~Casas, F.~Quevedo and E.~Roulet,
Phys.\ Lett.\ B {\bf 318}, 447 (1993)  [hep-ph/9308325];
  T.~Banks, D.~B.~Kaplan and A.~E.~Nelson,
  Phys.\ Rev.\ D {\bf 49}, 779 (1994)
  [hep-ph/9308292].


\bibitem{Berera:1995ie} 
  A.~Berera,
  Phys.\ Rev.\ Lett.\  {\bf 75}, 3218 (1995)
  [astro-ph/9509049];
  A.~Berera, I.~G.~Moss and R.~O.~Ramos,
  Rept.\ Prog.\ Phys.\  {\bf 72}, 026901 (2009)
  [arXiv:0808.1855 [hep-ph]];
  M.~Bastero-Gil and A.~Berera,
  Int.\ J.\ Mod.\ Phys.\ A {\bf 24}, 2207 (2009)
  [arXiv:0902.0521 [hep-ph]];

\bibitem{Yokoyama:2004pf} 
  J.~'i.~Yokoyama,
  Phys.\ Rev.\ D {\bf 70}, 103511 (2004)
  [hep-ph/0406072];
  J.~'i.~Yokoyama,
  Phys.\ Lett.\ B {\bf 635}, 66 (2006)
  [hep-ph/0510091].
  
  
\bibitem{Drewes:2010pf} 
  M.~Drewes,
  arXiv:1012.5380 [hep-th];
  M.~Drewes and J.~UKang,
  arXiv:1305.0267 [hep-ph].
  
  
\bibitem{Kofman:1994rk} 
  L.~Kofman, A.~D.~Linde and A.~A.~Starobinsky,
  Phys.\ Rev.\ Lett.\  {\bf 73}, 3195 (1994)
  [hep-th/9405187];
  Phys.\ Rev.\ D {\bf 56}, 3258 (1997)
  [hep-ph/9704452].
  
  
\bibitem{Mukaida:2012qn} 
  K.~Mukaida and K.~Nakayama,
  JCAP {\bf 1301}, 017 (2013)
  [arXiv:1208.3399 [hep-ph]].
  
\bibitem{Mukaida:2012bz} 
  K.~Mukaida and K.~Nakayama,
  JCAP {\bf 1303}, 002 (2013)
  [arXiv:1212.4985 [hep-ph]].
  

\bibitem{Silveira:1985rk}  
  V.~Silveira and A.~Zee,
  Phys.\ Lett.\ B {\bf 161}, 136 (1985);
  J.~McDonald,
  Phys.\ Rev.\ D {\bf 50}, 3637 (1994)
  [hep-ph/0702143 [HEP-PH]].
  
For recent analysis, see for instance:
  J.~M.~Cline, K.~Kainulainen, P.~Scott and C.~Weniger,
  arXiv:1306.4710 [hep-ph].

\bibitem{Okada:2010jd} 
  N.~Okada and Q.~Shafi,
  Phys.\ Rev.\ D {\bf 84}, 043533 (2011)
  [arXiv:1007.1672 [hep-ph]].
  
\bibitem{Enqvist:2012tc} 
  K.~Enqvist, D.~G.~Figueroa and R.~N.~Lerner,
  JCAP {\bf 1301}, 040 (2013)
  [arXiv:1211.5028 [astro-ph.CO]];
  K.~Enqvist, R.~N.~Lerner and S.~Rusak,
  arXiv:1308.3321 [astro-ph.CO].
  


\bibitem{Moroi:2013tea} 
  T.~Moroi, K.~Mukaida, K.~Nakayama and M.~Takimoto,
  JHEP {\bf 1306}, 040 (2013)
  [arXiv:1304.6597 [hep-ph]].
      

\bibitem{Dolan:1973qd} 
  L.~Dolan and R.~Jackiw,
  Phys.\ Rev.\ D {\bf 9}, 3320 (1974).
  
\bibitem{Anisimov:2000wx} 
  A.~Anisimov and M.~Dine,
  Nucl.\ Phys.\ B {\bf 619}, 729 (2001)
  [hep-ph/0008058].
  
\bibitem{Felder:1998vq} 
  G.~N.~Felder, L.~Kofman and A.~D.~Linde,
  Phys.\ Rev.\ D {\bf 59}, 123523 (1999)
  [hep-ph/9812289].
  
\bibitem{Kurkela:2011ti} 
  A.~Kurkela and G.~D.~Moore,
  JHEP {\bf 1112}, 044 (2011)
  [arXiv:1107.5050 [hep-ph]].
  

\bibitem{Bodeker:2006ij} 
  D.~Bodeker,
  JCAP {\bf 0606}, 027 (2006)
  [hep-ph/0605030];
  M.~Laine,
  Prog.\ Theor.\ Phys.\ Suppl.\  {\bf 186}, 404 (2010)
  [arXiv:1007.2590 [hep-ph]].
  
\bibitem{Moroi:2012vu} 
  T.~Moroi and M.~Takimoto,
  Phys.\ Lett.\ B {\bf 718}, 105 (2012)
  [arXiv:1207.4858 [hep-ph]].
  
  

\bibitem{Kadanoff:1962}
L.~P.~Kadanoff and G.~Baym, {\em ``Quantum Statistical Mechanics,''}
\newblock Benjamin New York (1962).


\bibitem{Baym:1961zz} 
  G.~Baym and L.~P.~Kadanoff,
  Phys.\ Rev.\  {\bf 124}, 287 (1961).

\bibitem{Cornwall:1974vz} 
  J.~M.~Cornwall, R.~Jackiw and E.~Tomboulis,
  Phys.\ Rev.\ D {\bf 10}, 2428 (1974).
  

\bibitem{Chou:1984es}
  K.~c.~Chou, Z.~b.~Su, B.~l.~Hao, L.~Yu,
  Phys.\ Rept.\  {\bf 118 } (1985)  1.
  
\bibitem{Berges:2004yj} 
  J.~Berges,
  AIP Conf.\ Proc.\  {\bf 739}, 3 (2005)
  [hep-ph/0409233].
  
\bibitem{calzetta2008nonequilibrium}  
  E.~A.~Calzetta and B.~L.~Hu,
  {\em "Nonequilibrium quantum field theory",}
  \newblock Cambridge University Press (2008).

\bibitem{Schwinger:1960qe}
  J.~S.~Schwinger,
  J.\ Math.\ Phys.\  {\bf 2 } (1961)  407-432;
  P.~M.~Bakshi and K.~T.~Mahanthappa,
  J.\ Math.\ Phys.\  {\bf 4 } (1963)  1, {\bf 4} (1963) 12;  
  L.~V.~Keldysh,
  Zh.\ Eksp.\ Teor.\ Fiz.\  {\bf 47 } (1964)  1515-1527.

\bibitem{Kubo:1957mj} 
  R.~Kubo,
  J.\ Phys.\ Soc.\ Jap.\  {\bf 12}, 570 (1957);
  P.~C.~Martin and J.~S.~Schwinger,
  Phys.\ Rev.\  {\bf 115}, 1342 (1959).
  
  
\bibitem{Aarts:2007ye} 
  G.~Aarts and A.~Tranberg,
  Phys.\ Rev.\ D {\bf 77}, 123521 (2008)
  [arXiv:0712.1120 [hep-ph]].
  
\bibitem{Tranberg:2008ae} 
  A.~Tranberg,
  JHEP {\bf 0811}, 037 (2008)
  [arXiv:0806.3158 [hep-ph]].
  
  
\bibitem{Berges:2002cz} 
  J.~Berges and J.~Serreau,
  Phys.\ Rev.\ Lett.\  {\bf 91}, 111601 (2003)
  [hep-ph/0208070];
  J.~Berges, A.~.Rothkopf and J.~Schmidt,
  Phys.\ Rev.\ Lett.\  {\bf 101}, 041603 (2008)
  [arXiv:0803.0131 [hep-ph]];
  J.~Berges, D.~Gelfand and J.~Pruschke,
  Phys.\ Rev.\ Lett.\  {\bf 107}, 061301 (2011)
  [arXiv:1012.4632 [hep-ph]];
  J.~Berges and D.~Sexty,
  Phys.\ Rev.\ Lett.\  {\bf 108}, 161601 (2012)
  [arXiv:1201.0687 [hep-ph]];
  J.~Berges, D.~Gelfand and D.~Sexty,
  arXiv:1308.2180 [hep-ph].
  
\bibitem{Garbrecht:2002pd} 
  B.~Garbrecht, T.~Prokopec and M.~G.~Schmidt,
  Eur.\ Phys.\ J.\ C {\bf 38}, 135 (2004)
  [hep-th/0211219].
  
\bibitem{Jeon:1994if} 
  S.~Jeon,
  Phys.\ Rev.\ D {\bf 52}, 3591 (1995)
  [hep-ph/9409250];
  S.~Jeon and L.~G.~Yaffe,
  Phys.\ Rev.\ D {\bf 53}, 5799 (1996)
  [hep-ph/9512263].
  
\bibitem{BasteroGil:2010pb} 
  M.~Bastero-Gil, A.~Berera and R.~O.~Ramos,
  JCAP {\bf 1109}, 033 (2011)
  [arXiv:1008.1929 [hep-ph]].
  
\bibitem{Kasuya:1996np} 
  S.~Kasuya and M.~Kawasaki,
  Phys.\ Lett.\ B {\bf 388}, 686 (1996)
  [hep-ph/9603317];
  M.~Hotta, I.~Joichi, S.~Matsumoto and M.~Yoshimura,
  Phys.\ Rev.\ D {\bf 55}, 4614 (1997)
  [hep-ph/9608374].
  
  
\bibitem{Calzetta:1986cq} 
  E.~Calzetta and B.~L.~Hu,
  Phys.\ Rev.\ D {\bf 37}, 2878 (1988).
  
\bibitem{Ivanov:1999tj} 
  Y.~.B.~Ivanov, J.~Knoll and D.~N.~Voskresensky,
  Nucl.\ Phys.\ A {\bf 672}, 313 (2000)
  [nucl-th/9905028].
  
\bibitem{Prokopec:2003pj} 
  T.~Prokopec, M.~G.~Schmidt and S.~Weinstock,
  Annals Phys.\  {\bf 314}, 208 (2004)
  [hep-ph/0312110];
  T.~Prokopec, M.~G.~Schmidt and S.~Weinstock,
  Annals Phys.\  {\bf 314}, 267 (2004)
  [hep-ph/0406140].
  
\bibitem{Boyanovsky:2004dj} 
  D.~Boyanovsky, K.~Davey and C.~M.~Ho,
  Phys.\ Rev.\ D {\bf 71}, 023523 (2005)
  [hep-ph/0411042].
  
\bibitem{Berges:2005md} 
  J.~Berges and S.~Borsanyi,
  Phys.\ Rev.\ D {\bf 74}, 045022 (2006)
  [hep-ph/0512155].
  
\bibitem{Hohenegger:2008zk} 
  A.~Hohenegger, A.~Kartavtsev and M.~Lindner,
  Phys.\ Rev.\ D {\bf 78}, 085027 (2008)
  [arXiv:0807.4551 [hep-ph]].
  
\bibitem{Anisimov:2008dz} 
  A.~Anisimov, W.~Buchmuller, M.~Drewes and S.~Mendizabal,
  Annals Phys.\  {\bf 324}, 1234 (2009)
  [arXiv:0812.1934 [hep-th]].
  
\bibitem{Garbrecht:2011xw} 
  B.~Garbrecht and M.~Garny,
  Annals Phys.\  {\bf 327}, 914 (2012)
  [arXiv:1108.3688 [hep-ph]].
  
\bibitem{Hamaguchi:2011jy} 
  K.~Hamaguchi, T.~Moroi and K.~Mukaida,
  JHEP {\bf 1201}, 083 (2012)
  [arXiv:1111.4594 [hep-ph]].
  
\bibitem{Drewes:2012qw} 
  M.~Drewes, S.~Mendizabal and C.~Weniger,
  Phys.\ Lett.\ B {\bf 718}, 1119 (2013)
  [arXiv:1202.1301 [hep-ph]].


  

\bibitem{Shtanov:1994ce} 
  Y.~Shtanov, J.~H.~Traschen and R.~H.~Brandenberger,
  Phys.\ Rev.\ D {\bf 51}, 5438 (1995)
  [hep-ph/9407247].


\end{thebibliography}
\end{document}